\documentclass[usegraphicx,usenatbib]{mn2e}
\usepackage{epsfig}
\usepackage{subfigure}
\usepackage{latexsym}
\usepackage{graphics}
\usepackage{graphicx}

\begin{document}

\title[Local and Global Variations of The Fine Structure Constant]{Local and
Global Variations of The Fine Structure Constant}
\author[D. F. Mota and J. D. Barrow]{David F. Mota$^{1,2}$\thanks{%
E-mail: D.F.Mota@damtp.cam.ac.uk} and John D. Barrow$^1$\thanks{%
E-mail: J.D.Barrow@damtp.cam.ac.uk} \\
%EndAName
$^1$ Department of Applied Mathematics and Theoretical Physics, Centre for
Mathematical Sciences, \\
University of Cambridge, Wilberforce Road, Cambridge CB3 0WA, UK\\
$^2$Astrophysics, Department of Physics, University of Oxford, Keble Road,
Oxford, OX1 3RH, UK}
\maketitle

\begin{abstract}
Using the BSBM varying-alpha theory, with dark matter dominated by magnetic
energy, and the spherical collapse model for cosmological structure
formation, we have studied the effects of the dark-energy equation of state
and the coupling of alpha to the matter fields on the space and time
evolution of alpha. We have compared its evolution inside virialised
overdensities with that in the cosmological background, using the standard ($%
\Lambda =0$) $CDM$ model of structure formation and the dark-energy
modification, $wCDM$. We find that, independently of the model of structure
formation one considers, there is always a difference between the value of
alpha in an overdensity and in the background. In a $SCDM$ model, this
difference is the same, independent of the virialisation redshift of the
overdense region. In the case of a $wCDM$ model, especially at low
redshifts, the difference depends on the time when virialisation occurs and
the equation of state of the dark energy. At high redshifts, when the $wCDM$
model becomes asymptotically equivalent to the $SCDM$ one, the difference is
constant. At low redshifts, when dark energy starts to dominate the
cosmological expansion, the difference between alpha in a cluster and in the
background grows. The inclusion of the effects of inhomogeneity leads
naturally to no observable local time variations of alpha on Earth and in
our Galaxy even though time variations can be significant on quasar scales.
The inclusion of the effects of inhomogeneous cosmological evolution are
necessary if terrestrial and solar-system bounds on the time variation of
the fine structure 'constant' are to be correctly compared with
extragalactic data.
\end{abstract}
\pagerange{\pageref{firstpage}--\pageref{lastpage}} \pubyear{2003}
\label{firstpage}
%
%\begin{pacs}
%98.80.-k  06.20.Jr
%\end{pacs}
%
\section{Introduction}
Studies of the time and spatial variations of the fine structure 'constant', 
$\alpha $, are motivated primarily by recent observations of small
variations in relativistic atomic structure in quasar absorption spectra %
\citep{murphy,murphy1,murphy2,murphy3}. Three data sets, containing
observations of the spectra of 128 quasars continue to suggest that the fine
structure 'constant', was smaller at redshifts $z=0.2-4.2$ than the current
terrestrial value $\alpha _{0}=7.29735308\times 10^{-3}$, with $\Delta
\alpha /\alpha \equiv \{\alpha (z)-\alpha _{0}\}/\alpha _{0}=-0.57\pm
0.11\times 10^{-5}$ \citep {murphylastlast}. If this interpretation of these
observations proves correct then there are important consequences for our
understanding of the forces of nature at low energies as well as for the
question of the links between couplings in higher dimensions and the
three-dimensional shadows that we observe \citep{book} . Any slow change in
the scale of the extra-dimensions would be revealed by measurable changes in
our four-dimensional \textquotedblleft constants\textquotedblright . Further
observational tests may help to provide independent tests of the quasar data %
\citep{darling}. Attempts to generalise the standard model by including
scalar fields which carry the space-time variation of the fine structure
'constant' could have important connections with the dark energy that is
currently accelerating the expansion of the universe \citep{bass,wet} and also
create potentially detectable violations of the weak equivalence principle %
\citep{wep,wep1,wep2,wep3}.

Several theories have been proposed to investigate the implications of a
varying fine structure 'constant'. Some based on grand unification theories %
\citep{lang,lang1,lang2,lang3,lang4,lang5,lang6,lang7}, some with
extra-dimensions \citep{carsten,carsten1,carsten2,dent,correia},  others in
four dimensions \citep{bek,bek1,bek2,bek3,bek4,bsbm,mota1,carsten3,ralf} and
even in two dimensional black holes \citep{vagenas}. All these models need
to satisfy the present constraints on the variations in $\alpha $. These
constraints can be divided into two main groups: local and
astro-cosmological. The local constraints derive from experiments in our
local bound gravitational system: the Oklo natural reactor $(z=0.14)$, where 
$\left\vert \frac{\Delta \alpha }{\alpha }\right\vert \leq 10^{-7}$ %
\citep{fuj,fuj1,fuj2,fuj3}. However, the use of
an equilibrium neutron spectrum in this analyses has recently been
criticised by \citep{lamoreaux}. More realistic
modelling leads to a best fit of the data to non-zero shift
$\Delta\alpha/\alpha = 4.5 \times 10^{-8}$. 
; the intra solar-system decay rate $^{187}\mathrm{%
Re}~\rightarrow ~^{187}\mathrm{Os}$, $(z=0.45)$, where $\left\vert \frac{%
\Delta \alpha }{\alpha }\right\vert \leq %\times
10^{-7}$ \citep{olive,iwanoto,olivere}; and the stability of terrestrial atomic clocks $(z=0)
$, where $\left\vert \frac{\dot{\alpha}}{\alpha }\right\vert <4.2\times
10^{-15}~\mathrm{yr}^{-1}$ \citep{marion}. Other limits arise from weak
equivalence experiments \citep{wep,wep1,wep2,wep3,wep4} but the limits they
provide are more model dependent. In addition to the the above-mentioned
positive signal from quasar absorption spectra, the astro-cosmological
constraints are: the cosmic microwave background radiation $(z=10^{3})$,
where $\left\vert \frac{\Delta \alpha }{\alpha }\right\vert <10^{-2}$ %
\citep{martins,martins1,martins2,steen,turner} and Big Bang nucleosynthesis $%
(z=10^{8}-10^{10})$, where $\left\vert \frac{\Delta \alpha }{\alpha }%
\right\vert \leq 2\times 10^{-2}$ \citep{martins,martins1,martins2}. Other
emission-line studies \citep{bah} give weak bounds of $\frac{\Delta \alpha }{%
\alpha }\leq -2\pm 1.2\times 10^{-4}$ which are comparable to the earlier
limits of refs. \citep{song,ivan,lev}.

There is a potential discrepancy between local and astro-cosmological
constraints; in particular, between the constraint of $\Delta \alpha /\alpha
\leq 10^{-7}$ at redshift $0\leq z\leq0.45$, coming from the $\beta $-decay in
meteoritic samples \citep{olive,olivere}, and the explicit variation in $\alpha $ of 
$\Delta \alpha /\alpha \approx 10^{-6}$ at $0.2\leq z \leq4.2$, coming from the
low-redshift end of the quasar absorption spectra %
\citep{murphylast,murphy,murphy1,murphy2,murphy3}. A successful theory of varying $%
\alpha $ needs to explain this difference. This is a challenge. Models which
use a very light scalar field, to drive variations in $\alpha $, need an
extreme fine tuning in order to satisfy the phenomenological constraints
coming from geochemical data (Oklo, $\beta $-decay), the present equivalent
principle tests, and the quasar absorption spectra, simultaneously %
\citep{dam}. The presence of the dark energy plays an important role in any
potential reconciliation because in simple scalar theories \citep{bsbm} any
variation of $\alpha $ turns off as the universe starts to accelerate.

A solution to this problem was proposed by us in \citep{mota3}, where
numerical simulations were shown to display behaviour for the evolution of $%
\alpha $ inside an overdensity that differs from that in the background
universe, during the formation of non-linear large-scale structures. All
previous studies of the variation of 'constants' in cosmology had neglected
the effects of inhomogeneity in the Universe and the fact that our local
observational constraints on varying constants are made within non-expanding
matter overdensities. We found that the fine structure 'constant' evolves
differently inside virialised clusters compared to the background universe.
Specifically, $\alpha $ becomes a effectively time-independent inside a
gravitationally bound overdensity after its virialisation. The fact that
local $\alpha $ values 'freeze in' at virialisation, means we would observe
no time variations in $\alpha $ on Earth, or elsewhere in our Galaxy, even
though time-variations in $\alpha $ might still be occurring on
extragalactic scales at a detectable level. For a typical galaxy cluster,
the value of $\alpha $ today will be the value of $\alpha $ at the
virialisation time of the cluster. Hence, the local constraints on time
variations in the fine structure 'constant', can easily give a value that is 
$10-100$ times smaller than is inferred on extra-galactic scales from quasar
absorption spectra.

The reason why the growth and gravitational binding of matter
inhomogeneities affects the evolution of $\alpha $ is simple. Any varying-$%
\alpha $ theory implies the existence of a scalar field carrying the
variations in $\alpha $. This field is coupled to some selection of the
matter fields, depending upon their interactions \citep{bass, gold,wet}. Due to
this coupling, any inhomogeneities of the latter will then affect the
evolution of $\alpha $. This effect is not only important in the non-linear
regime of large-scale structure formation \citep{otoole}. Even in the linear
regime of cosmological perturbations, when these are small, $\frac{\delta
\alpha }{\alpha }$ grows and tracks $\frac{\delta \rho _{m}}{\rho _{m}}%
\propto t^{2/3}$ during the dust-dominated era on scales smaller than the
Hubble radius \citep{mota2}. It is therefore important to study the
cosmological evolution of the fine structure 'constant' taking into account
a more realistic universe, where matter inhomogeneities grow and lead to the
formation of bounded objects.

The dependence of the fine structure constant on the density of the matter
fields leads to spatial variations of $\alpha $. Spatial variations imply
the existence of 'fifth force' effects. The fifth force induces an anomalous
acceleration which depends on the material composition of the test particle,
and so violates the weak equivalence principle (WEP). This is a general
feature of any varying-$\alpha $ theory but he magnitude of the WEP
violations is model dependent. In the case of the
Bekenstein-Sandvik-Barrow-Magueijo (BSBM) model, such violations are within
the current WEP experiments \citep{otoole1}. In particular, it was shown in %
\citep{wep4} that spatial variations of $\alpha $ in the vicinity of compact
massive objects may well be within an order of magnitude below the existing
experimental bounds (see appendix \ref{wepappendix}).

There are two main features which affect the evolution of the fine structure
'constant' in a universe where large scale structures are formed. The most
obvious, is the coupling between the scalar field, which drives variations
in $\alpha $, and the matter fields. %The value of the coupling will
%depend on the nature of the matter fields. 
The second is the dependence of non-linear models of structure formation on
the equation of state of the universe, and in particular that of the dark
energy \citep{lahav,lahav1}. In this paper we will study the dependence of $%
\alpha $ on these two quantities in the BSBM varying-$\alpha $ model. In
this theory, variations in $\alpha $ are driven by the
electromagnetically-coupled matter fields and the effects of inhomogeneity
are large. In the next section we briefly summarise the BSBM model \citep{%
bsbm} and the spherical collapse model for the development of non-linear
cosmological inhomogeneities. In section three, we investigate the
dependence of the fine structure 'constant' on the equation of state of the
dark energy. In section four, we show how spatial variations in $\alpha $
may occur due to possible spatial variations in the coupling of $\alpha $ to
the matter fields. We summarise our conclusions in section five.

\section{The BSBM Theory and the Spherical-Infall Model}

\subsection{The Background}

We will study space-time variations of $\alpha $ in the BSBM theory %
\citep{bsbm}, which assumes that the total action is given by: \ 
\begin{equation}
S=\int d^{4}x\sqrt{-g}\left( \mathcal{L}_{g}+\mathcal{L}_{matter}+\mathcal{L}%
_{\psi }+\mathcal{L}_{em}e^{-2\psi }\right)   \label{S}
\end{equation}%
In this varying-$\alpha $ theory, the quantities $c$ and $\hbar $ are taken
to be constant, while $e$ varies as a function of a real scalar field $\psi
(x^{\beta }),$ with $e=e_{0}e^{\psi }$, $\mathcal{L}_{\psi }={\frac{\omega }{%
2}}\partial _{\mu }\psi \partial ^{\mu }\psi $, $\omega $ is a coupling
constant, and $\mathcal{L}_{em}=-\frac{1}{4}f_{\mu \nu }f^{\mu \nu }$; $%
\mathcal{L}_{matter}$ is the Lagrangian of the matter fields. The
gravitational Lagrangian is as usual $\mathcal{L}_{g}=-\frac{1}{16\pi G}R$,
with $R$ the Ricci curvature scalar, and we have defined an auxiliary gauge
potential by $a_{\mu }=\epsilon A_{\mu }$ and a new Maxwell field tensor by $%
f_{\mu \nu }=\epsilon F_{\mu \nu }=\partial _{\mu }a_{\nu }-\partial _{\nu
}a_{\mu }$, so the covariant derivative takes the usual form, $D_{\mu
}=\partial _{\mu }+ie_{0}a_{\mu }$. The fine structure 'constant' is then
given by $\alpha \equiv \alpha _{0}e^{2\psi }$ with $\alpha _{0}$ the
present value measured on Earth today.

The background universe will be described by a flat, homogeneous and
isotropic Friedmann metric with expansion scale factor $a(t)$. The universe
contains pressure-free matter, of density $\rho _{m}$ $\propto a^{-3}$ and a
dark-energy fluid with a constant equation of state parameter, $w_{\phi
}=p_{\phi }/\rho _{\phi }$, and an energy-density $\rho _{\phi }\propto
a^{-3(1+w_{\phi })}$. In the case where dark energy is the cosmological
constant $\Lambda $, $\rho _{\phi }\equiv \rho _{\Lambda }\equiv $ $\Lambda
/(8\pi G)$ and $w_{\phi }=-1$. Varying the total Lagrangian, we obtain the
Friedmann equation ($\hbar =c\equiv 1$): 
\begin{equation}
H^{2}=\frac{8\pi G}{3}\left( \rho _{m}\left( 1+\left\vert \zeta \right\vert
e^{-2\psi }\right) \ +\rho _{\psi }+\rho _{\phi }\right) ,  \label{fried1}
\end{equation}%
where $H\equiv \dot{a}/a$ is the Hubble rate, $\rho _{\psi }=\frac{\omega }{2%
}\dot{\psi}^{2},$ and $\zeta =\mathcal{L}_{em}/\rho _{m}$ is the fraction of
the matter which carries electric or magnetic charges. 
The value  (and sign) of $\zeta $ for baryonic and dark matter has been disputed \citep
{wep,olive,bsbm}. It is the difference between the percentage of mass in
electrostatic and magnetostatic forms. As explained in \citep{bsbm}, we can
at most \textit{estimate} this quantity for neutrons and protons, with $%
\zeta _{n}\approx \zeta _{p}\sim 10^{-4}$. We may expect that for baryonic
matter $\zeta \sim 10^{-4}$, with composition-dependent variations of the
same order. The value of $\zeta $ for the dark matter, for all we know,
could be anything between -1 and 1. Superconducting cosmic strings, or
magnetic monopoles, display a \textit{negative} $\zeta $, unlike more
conventional dark matter.  It is clear that the only way to obtain a
cosmologically increasing $\alpha $ in BSBM is with $\zeta <0$, i.e with
unusual dark matter, in which magnetic energy dominates over electrostatic
energy.
 
The term $\left\vert \zeta \right\vert e^{-2\psi }$ represents
an average\footnote{%
Each species in the Friedmann equation may have associated with it a
different value of $\zeta$. However, since we are interested in the
large-scale cosmological behaviour we have averaged all matter
contributions  into a mean effective 'cosmological'
value of $\zeta$ (see appendix \ref{bsbmeq}).} of the (always
positive) energy density contribution from the non-relativistic matter which
interacts electromagnetically.

The scalar-field evolution equation is 
\begin{equation}
\ddot{\psi}+3H\dot{\psi}=-\frac{2}{\omega }\ e^{{-2\psi }}\zeta \rho _{m}.
\label{psidot}
\end{equation}

In this article, we consider that the scalar field responsible for the
variations of $\alpha$ is coupled to dark matter and to baryons. Both
species are included in the $\zeta\rho_m$ terms. The coupling to dark matter
appears in general in any dilaton type theory, like string theory, where a
scalar field is coupled to all the terms in the Lagrangian, but not
necessarily in the same way. In the case of BSBM models, the coupling is
motivated by certain classes of dark-matter models and their supersymmetric
versions \cite{wep1}. The coupling between $\alpha$ and dark-matter could
permit dark-matter-photon interactions of which would be mediated by the
scalar field. Due to the ``invisible'' nature of dark matter, its coupling
to $\alpha$ is tightly constrained. In the case of BSBM models several
constraints were imposed on $\zeta$ in \citep{wep1}. Other motivations and
consequences of this coupling are investigated in \cite{celine}, where
dark-matter-photon interactions are constrained using the CMBR anisotropies
and the matter power spectrum. In \cite{celine} the cross
section associated to the dark-matter -
photon interaction is constraint to be
$\sigma_{\gamma-DM}/m_{DM}\leq 10^{-6}\sigma_{th}/(100\, GeV)
\approx6\times 10^{-33} cm^2\, GeV^{-1}
$, where $\sigma_{\gamma-DM}$ is the
dark-matter-photon cross section, $m_{DM}$ is the dark-matter particle
mass and $\sigma_{th}$ is the Thomson
cross section. Since in our case $\sigma_{th}\sim\alpha^2\sim\alpha_0^2
e^{4\psi}$, 
then an approximate constraint to
the coupling between the scalar field and dark-mater can be set
imposing that
$\sigma_{\psi-DM}\sigma_{\psi-\gamma}\leq\sigma_{\gamma-DM}$, where
$\sigma_{\psi-DM}\sim \zeta^2 e^{-4\psi}/\omega^2$ and 
$\sigma_{\psi-\gamma}\sim e^{-4\psi}$. 
Roughly
giving 
$\vert\zeta \vert\leq 10^{-3}\alpha_0\sim 10^{-6}$, when assuming the ``usual''
$m_{DM}\sim100\, GeV$. In this article we will use $\vert\zeta \vert\sim
10^{-8}$ as a reference value. 
In reality, as we
show in section 4, the cosmological evolution of the fine structure
constant is independent of value of $\vert\zeta \vert$ due to a
degeneracy between the initial condition for $\psi$ and the
coupling to the matter fields. 
The independence of our results with respect to the value of
$\zeta/\omega$ give us the freedom to make the coupling to dark matter as
small as desired. For instance, we are free to make the 
cross section photon - scalar field
- dark matter smaller enough to satisfy all the constraints coming from
the ``invisible'' nature of dark-matter. In a similar way, we are free to
constraint the value of $\zeta/\omega$ in order to avoid problems
associated with
the so called self-interacting dark matter, which could
radiate and cool, due to its 'magnetic' nature. All these features
could be combined to constraint the value of $\zeta$. However that
would be a lengthy and a highly model-dependent analysis, and will be
considered by the authors elsewhere.

\subsection{The Density Inhomogeneities}

In order to study the behaviour of the fine structure 'constant' inside
overdensities we will use the spherical-infall model \citep{padmanaban}.
This will describe how the field in the overdensity breaks away from the
field in background expansion. The overdense sphere behaves like a spatially
closed sub-universe. The density perturbations need not to be uniform within
the sphere: any spherically symmetric perturbation will evolve within a
given radius in the same way as a uniform sphere containing the same amount
of mass. Similar results could be obtained by performing the analysis of the
BSBM theory using a spherically symmetric Tolman-Bondi metric for the
background universe with account taken for the existence of the pressure
contributed by the dark energy and the $\psi $ field. In what follows,
therefore, density refers to \textit{mean} density inside a given sphere.

Consider a spherical perturbation with constant internal density which, at
an initial time, has an amplitude $\delta _{i}>0$ and $|\delta _{i}|\ll 1$. 
%For the case of a flat
%matter-dominated FRW universe it is possible to find an exact solution for the
%evolution of the overdensity, although that is not the case for more
%complex cases, like ours, and a numerical simulation is
%required. Nevertheless the general behaviour of the overdensity will
%be similar. 
At early times the sphere expands along with the background. For a
sufficiently large $\delta _{i},$ gravity prevents the sphere from
expanding. Three characteristic phases can then be identified. \textit{%
Turnaround}: the sphere breaks away from the general expansion and reaches a
maximum radius. \textit{Collapse}: if only gravity is significant, the
sphere will then collapse towards a central singularity where the densities
of the matter fields would formally go to infinity. In practice, pressure
and dissipative physics intervene well before this singularity is reached
and convert the kinetic energy of collapse into random motions. \textit{%
Virialisation}: dynamical equilibrium is reached and the system becomes
stationary: there are no more time variations in the radius of the system, $%
R $, or in its energy components. This phase determines the final pattern of
variations in $\alpha ,$ which becomes a constant inside the virialised
region, otherwise the system would be unable to virialise \citep{mota3}.
Meanwhile, depending upon the equation of state of the dominant matter
field, $\alpha $ can continue to change in the cosmological background. This
behaviour naturally creates a situation where time variation of $\alpha $ on
large cosmological scales is accompanied by unchanging behaviour locally
within galaxies and our solar system. It would explain the apparent
discrepancy between the results of the quasars absorption spectra
observations and the $\beta $-decay rate deductions from meteorite data %
\citep{olive}.

The evolution of a spherical overdense patch of scale radius $R(t)$ is given
by the Friedmann acceleration equation: 
\begin{equation}
\frac{\ddot{R}}{R}=-\frac{4\pi G}{3}\left( \rho _{cdm}\left( 1+\left\vert
\zeta \right\vert e^{-2\psi _{c}}\right) +4\rho _{\psi _{c}}+(1+3w_{\phi
_{c}})\rho _{\phi _{c}}\right) ,  \label{rcluster1}
\end{equation}%
where $\rho _{cdm}$ is the density of cold dark matter in the cluster, $\rho
_{\phi _{c}}$ is the energy density of the dark energy inside the cluster
and $\rho _{\psi _{c}}\equiv \frac{\omega }{2}\dot{\psi}_{c}^{2}$, where $%
\psi _{c}$ represents the scalar field inside the overdensity. We have also
used the equations of state $p_{\psi _{c}}=\rho _{\psi _{c}}$, $p_{cdm}=0$
and $p_{\phi _{c}}=w_{\phi _{c}}\rho _{\phi _{c}}$.

In the cluster, the evolutions of $\psi _{c}$, $\rho _{cdm}$ and $\rho
_{\phi _{c}}$ are given by 
\begin{eqnarray}
\ddot{\psi _{c}}+3\frac{\dot{R}}{R}\dot{\psi _{c}} &=&-\frac{2}{\omega }%
e^{-2\psi _{c}}\zeta \rho _{cdm},  \label{psidotcluster1} \\
\dot{\rho}_{cdm} &=&-3\frac{\dot{R}}{R}\rho _{cdm},  \label{rhocluster1} \\
\dot{\rho}_{\phi _{c}} &=&-3\frac{\dot{R}}{R}\left( 1+w_{\phi _{c}}\right)
\rho _{\phi _{c}}.  \label{dark1}
\end{eqnarray}

We will evolve the spherical overdensity from high redshift until its
virialisation occurs. According to the virial theorem, equilibrium will be
reached when $T=\frac{1}{2}R\frac{\partial U}{\partial R}$; $T$, is the
average total kinetic energy, and $U$ is the average total potential energy
in the sphere. Note that we obtain the usual $T=\frac{n}{2}U$ condition, for
any potential with a power-law form ($U\propto R^{n}$), which includes our
case. It is useful to write the condition for virialisation to occur in
terms of the potential energies associated the different components of the
overdensity. The potential energy for a given component $^{\prime }x^{\prime
}$ can be calculated from its general form in a spherical region \citep{%
landaubook}: 
\begin{eqnarray}
U_{x} &=&2\pi \int_{0}^{R}\rho _{tot}\phi _{x}r^{2}dr, \\
\phi _{x}\left( r\right) &=&-2\pi G\rho _{x}\left( R^{2}-\frac{r^{2}}{3}%
\right) ,  \label{generalpotential}
\end{eqnarray}%
where $\rho _{tot}$ is the total energy density inside the sphere, $\phi
_{x} $ is the gravitational potential due to the $\rho _{x}$ density
component.

In the case of a $\Lambda CDM$ model, the potential energies inside the
cluster are: 
\begin{eqnarray}
U_{G} &=&-\frac{3}{5}GM^{2}R^{-1}, \\
U_{\Lambda } &=&-\frac{4}{5}\pi G\rho _{\Lambda }MR^{2}, \\
U_{\psi _{c}} &=&-\frac{3}{5}GMM_{\psi _{c}}R^{-4},  \label{potential}
\end{eqnarray}%
where $U_{G}$ is the potential energy associated with the uniform spherical
overdensity, $U_{\Lambda }$ is the potential associated with $\Lambda ,$ and 
$U_{\psi _{c}}$ is the potential associated with $\psi _{c}.$ $%
M=M_{cdm}+M_{\psi _{c}}$ is the cluster mass, with 
\begin{eqnarray}
M_{cdm} &=&\frac{4\pi }{3}\rho _{cdm}(1+|\zeta |e^{-2\psi _{c}})R^{3}, \\
M_{\psi _{c}} &=&\frac{4\pi }{3}\rho _{\psi _{c}}R^{6}.  \label{masses}
\end{eqnarray}

The virial theorem will be satisfied when 
\begin{equation}
T_{vir}=-\frac{1}{2}U_{G}+U_{\Lambda }-2U_{\psi _{c}},  \label{virial}
\end{equation}%
where $T_{vir}$ $=\frac{1}{2}M\bar{v}_{vir}^{2}$ is the total kinetic energy
at virialisation and $\bar{v}_{vir}^{2}$ is the mean-square velocity of the
components of the cluster.

Using the virial theorem (\ref{virial}) and energy conservation at the
turnaround and cluster virialisation times, we obtain an equilibrium
condition only in terms of the potential energies: 
\begin{equation}
\frac{1}{2}U_{G}(z_{v})+2U_{\Lambda }(z_{v})-U_{\psi
_{c}}(z_{v})=U_{G}(z_{ta})+U_{\Lambda }(z_{ta})+U_{\psi _{c}}(z_{ta}),
\label{virialcond}
\end{equation}%
where $z_{v}$ is the redshift of virialisation and $z_{ta}$ is the redshift
at the turnaround of the over-density at its maximum radius, when $R=R_{max}$
and $\dot{R}\equiv 0$. In the case where $\zeta =0$ and $\psi =\dot{\psi}=0$
we reduce to the usual virialisation condition for $\Lambda CDM$ models with
no variation of $\alpha $ \citep{lahav,lahav1}. The generalisation from the
cosmological constant case to a dark-energy fluid with a general equation of
state $w_{\phi }$ is straightforward. One just needs to note that the
potential energy associated to the dark-energy fluid is 
\[
U_{\phi _{c}}=-\frac{3}{5}GM_{\phi _{c}}MR^{2-3\left( 1+w_{\phi _{c}}\right)
}, 
\]%
where $M_{\phi _{c}}$ is mass associated of that fluid, which is given by 
\[
M_{\phi _{c}}=\frac{4\pi }{3}\rho _{\phi _{c}}R^{3\left( 1+w_{\phi
_{c}}\right) }. 
\]

%
%The inclusion of a varying $\alpha $ at a level consistent with observation 
% \citep{murphy,murphylast}, ($\dot{\alpha}/\alpha _{0}\sim
%10^{-6}H$), does not affect 
%the overall expansion of the universe up to logarithmic corrections,
%if we are far from the initial singularity   \citep{mota1}. The
%reason for that is the    negligible contribution of $\psi$  to the energy content
%of the universe  \citep{bsbm}.
%In a similar way, the dynamical collapse of the
%overdense regions is also not affected: The energy density associated
%with $\psi_c $ is always a negligible contribution to the energy content
%of cluster, see Figure \ref{cluster}. 
%Even the kinetic energy, $\rho_{\psi_c}=\frac{\omega}{2}\dot\psi^2$, which starts to
%grow in the final stages of the cluster evolution will be
%negligible, see Figure \ref{cluster}.
%
%\begin{figure}[htbp!]
%\includegraphics[height=6.7cm,width=8cm]{cluster.eps}
%,height=6.7cm,width=8cm}
%\caption{{\protect {\it Evolution of  $\log \rho_{cdm}$ (solid lines) , 
%$\log \rho_{\psi_c}$ (dash-dotted line), $\log \rho_{\lambda}$ (dotted
%line) and  $\log \rho_{\alpha}$ (dashed line)  
%inside an overdensity, as a function of $\log (a)$, in a $\Lambda CDM$
%model. Where $\rho_{\alpha}\equiv \rho_{\psi_c} + \vert\zeta\vert \rho_{cdm} e^{-2\psi_c}$. }}}
%\label{cluster}
%\end{figure}
%
%

\subsection{Setting the Initial Conditions}

The behaviour of the fine structure 'constant' during the evolution of a
cluster can now be obtained numerically by evolving the background Friedmann
equations (\ref{fried1}) and (\ref{psidot}) together with the cluster
evolution equations (\ref{rcluster1})-(\ref{dark1}) until the virialisation
condition holds. In order to satisfy the constraints imposed by the
observations, we need to set up initial conditions for the evolution. Since
the Earth is now inside a virialised overdense region, the initial condition
for $\psi $ is chosen so as to obtain our measured laboratory value of $%
\alpha $ at virialisation, $\alpha _{c}(z_{v})\equiv \alpha _{v}=\alpha _{0}$%
. But, since the redshift at which our cluster has virialised is uncertain,
we will choose a representative example where virialisation occurs over the
range $0<z_{v}<5$. This is just for illustrative purposes, since in reality,
the initial condition for $\psi $ needs to be fixed only once, for our
Galaxy. Hence, $\alpha $ in other clusters will have a lower or higher value
(with respect to $\alpha _{0}$) depending on their $z_{v}$ values \citep{%
mota3}.

After we have set $z_{v}$ for Earth, another constraint we need to satisfy
is given by the quasar observations \citep{murphylast}. This means that when
comparing the value of the fine structure 'constant' on Earth, at its
virialisation, $\alpha _{v}=\alpha _{0}$, with the value of the fine
structure 'constant' of another region at some given redshift in the range
accessed by the quasar spectra, $3.5\geq z\geq 0.5$, we need to obtain $%
\Delta \alpha /\alpha \equiv (\alpha (z)-\alpha _{v})/\alpha _{v}\approx
-5.4\times 10^{-6}$. This raises the question as to the location of the
clouds where the quasar absorption lines are formed: are they in a region
which should be considered as part of the background or in an overdensity
with somewhat lower contrast than exists in our Galaxy? Unfortunately, this
question cannot be answered because we do not know the density of the
clouds, only the column density. Nevertheless, these clouds are much less
dense than the solar system. Because of this, it is a very good
approximation to assume that the clouds possess the background density.

Thus, the initial conditions for $\psi $ are chosen so as to obtain our
measured laboratory value of $\alpha $ at virialisation $\alpha
_{c}(z_{v})=\alpha _{0}$ and to match the latest observations \citep{%
murphylast} for background regions at $3.5\geq |z-z_{v}|\geq 0.5$.

Here we may also wonder about possible measurements of $\alpha $ in
Lyman-alpha systems and whether similar considerations should be applied
because of local variations in density compared to this in the solar
system. Assuming that the density of the Lyman-alpha systems are
much smaller than the Earth, we would expect to find a difference in
$\alpha$ with respect to the value measured on Earth. This difference
could even be of the same order as the one found in the quasar
spectra. However, this is dependent on the density of the clouds and redshift we are
making the measurements.  
Unfortunately we do not know the local density, only the column
density along the line of sight. In order to make some numerical predictions, we would
need to create a model populated with clouds of different density but there
are too many variables to make a reliable estimate of the effects at
present. In the future it may be possible,
with the accumulation of very large archives of data, to exploit the known
differences in column density in the different types of system where the
absorption lines are forms to search for correlations with column density.
The damped Lyman-alpha and Lyman-limit systems in the quasar
absorption-system data sets have column densities of order or exceeding $2
\times 10^{17} cm^{-1}$ and $2 \times 10^{20} cm^{-1}$ respectively.
Again, rigorous exploitation of any perceived systematic trend in alpha
between systems of different column density would require
the  construction of a HI density model in the vicinity of the absorption
 system.  

%This is an interesting direction for future investigation.

\section{The Dependence on the Dark-Energy Equation of Sate}

Non-linear models of structure formation will present different features
depending on the equation of state of the universe \citep{lahav,lahav1}. The
main difference is the way the parameter $\Delta _{c}=\rho
_{cdm}(z_{c})/\rho _{b}(z_{v})$ evolves with the redshift. The evolution of $%
\Delta _{c}$ depends on the equation of state of the dark-energy component
which dominates the expansion dynamics. For instance, in a $\Lambda CDM$
model, the density contrast, $\Delta _{c}$ increases as the redshift
decreases. At high redshifts, the density contrast at virialisation becomes
asymptotically constant in standard ($\Lambda =0$) $CDM$, with $\Delta
_{c}\approx 178$ at collapse or $\Delta _{c}\approx 148$ at virialisation.
This behaviour is common to other dark-energy models of structure formation (%
$wCDM$), where the major difference is in the magnitude and the rate of
change of $\Delta _{c}$ at low redshifts.

The local value of the fine structure 'constant' will be a function of the
redshift and will be dependent on the density of the region of the universe
we measure it, according to whether it is in the background or an
overdensity \citep{mota3}. The density contrast of the virialised clusters
depends on the dark-energy equation of state parameter, $w_{\phi }$ . Hence,
the evolution of $\alpha $ will be dependent on $w_{\phi }$ as well.

What difference we would expect to see in $\alpha $ if we compared two bound
systems, like two clusters of galaxies? And how does the difference depend
on the cosmological model of structure formation? These questions can be
answered by looking at the time and spatial variations of $\alpha $ at the
time of virialisation. The space variations will be tracked using a
'spatial' density contrast, 
\begin{equation}
\frac{\delta \alpha }{\alpha }\equiv \frac{\alpha _{c}-\alpha _{b}}{\alpha
_{b}}  \label{deltaspatialalpha}
\end{equation}%
which is computed at virialisation (where $\alpha _{c}(z=z_{v})=\alpha _{v}$%
). We assume there are no changes after this time.

Since the main dependence of $\alpha $ is on the density of the clusters and
the redshift of virialisation, we will only study dark-energy models where $%
w_{\phi }$ is a constant. Any effect contributed by a time-varying equation
of state should be negligible, since the important feature is the average
equation of state of the universe. This may not be the case for the models
where the scalar field responsible for the variations in $\alpha $ is
coupled to dark energy \citep{bass,copeland,gold}.

In order to have a qualitative behaviour of the evolution of the fine
structure 'constant', at the virialisation of an overdensity, we will then
compare the standard Cold Dark Matter model, $SCDM$, to the dark-energy Cold
Dark Matter, $wCDM$, models. In particular, we will examine the
representative cases of $w_{\phi }=-1,-0.8,-0.6$. All models will be
normalised to have $\alpha _{v}=\alpha _{0}$ at $z=0$ and to satisfy the
quasar observations \citep{murphylast}, as discussed above. This
normalisation, although unrealistic (Earth did not virialise today), give us
some indication of the dependence of the time and spatial evolution in $%
\alpha $ on the different models. In reality, this approximation will not
affect the order of magnitude of the spatial and time variations in $\alpha $
for the cases of virialisation at low redshift \citep{mota3}.

\subsection{Time-shifts and the evolution of $\protect\alpha $}

\begin{figure}%[tbp]
\epsfig{file=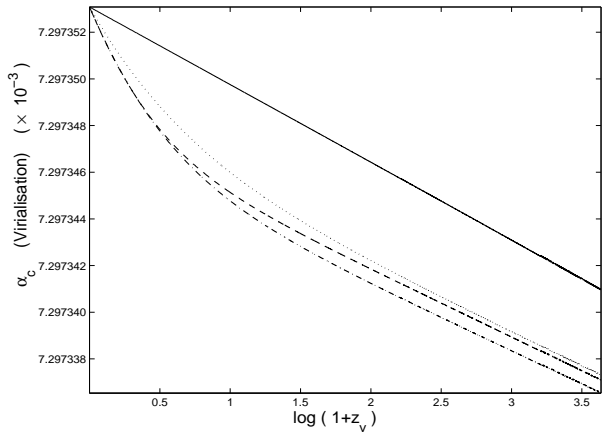,height=6cm,width=8cm}
\caption{\textit{Value of $\protect\alpha $ inside clusters as a function of 
$\log (1+z_{v})$, the epoch of virialisation. Solid line corresponds to the
standard ($\Lambda =0$) $CDM$, dashed line is the $\Lambda CDM$,
dashed-dotted corresponds to a $wCDM$ model with $w_{\protect\phi }=-0.8$,
dotted-line corresponds to a $wCDM$ model with $w_{\protect\phi }=-0.6$. In
all the models, the initial condition were set in order to have $\protect%
\alpha _{c}(z=0)=\protect\alpha _{0}$ and to satisfy $\Delta \protect\alpha /%
\protect\alpha \approx -5.4\times 10^{-6}$ at $3.5\geq |z-z_{v}|\geq 0.5$.}}
\label{modelsvirial}
\end{figure}

The final value of $\alpha $ inside virialised overdensities and its
evolution in the background is shown in Figures \ref{modelsvirial} and \ref%
{modelsvirialback}, respectively. From these plots, a feature common to all
the models stands out: the fine structure 'constant' in the background
regions has a lower value than inside the virialised overdensities. Also,
its eventual local value depends on the redshift at which the overdensity
virialises. 
\begin{figure}%[tbp]
\epsfig{file=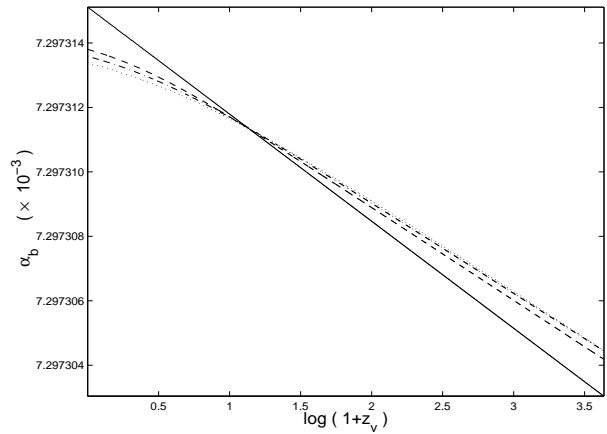,height=6cm,width=8cm}
\caption{\textit{The value of $\protect\alpha $ in the background\ universe
at the epoch of cluster virialisation, $\log (1+z_{v})$. The solid line
corresponds to the standard ($\Lambda =0$) $CDM$, dashed line is the $%
\Lambda CDM$, dashed-dotted line corresponds to a $wCDM$ model with $w_{%
\protect\phi }=-0.8$, dotted line corresponds to a $wCDM$ model with $w_{%
\protect\phi }=-0.6$. In all models, the initial condition were set in order
to have $\protect\alpha _{c}(z=0)=\protect\alpha _{0}$ and to satisfy $%
\Delta \protect\alpha /\protect\alpha \approx -5.4\times 10^{-6}$ at $%
3.5\geq |z-z_{v}|\geq 0.5$.}}
\label{modelsvirialback}
\end{figure}

As expected, the equation of state of the dark energy affects the evolution
of $\alpha ,$ both in the overdensities and the background. A major
difference arises if we compare the $SCDM$ and $wCDM$ models. In a $SCDM$
model, the fine structure 'constant' is always a growing function, both in
the background and inside the overdensities, and the growth rate is almost
constant. In a $wCDM$ model, the evolution of $\alpha $ will depend strongly
on whether one is inside a cluster or in the background. In the $wCDM$
background, $\alpha _{b}$ becomes constant (independent of the redshift) as
the universe enters the phase of accelerated expansion. Inside the clusters, 
$\alpha _{v}$ will always grow and its value now will depend on the redshift
at which virialisation occurred. The cumulative effect of this growth
increases as we consider overdensities which virialise at increasingly lower
redshift.

These differences arise due to the dependence of the fine structure
'constant' on the equation of state of the universe and the density of the
regions we are measuring it. In a $SCDM$ model, we will always live in a
dust-dominated era. The fine structure 'constant' will then be an
ever-increasing logarithmic function of time, $\alpha \propto \ln (t)$ \citep%
{bsbm,mota1}. The growth rates of $\alpha _{b}$ and $\alpha _{c}$ will be
constant, since $\Delta _{c}$ is independent of the redshift in a $SCDM$
model. In a $wCDM$ model, dark energy plays an important role at low
redshifts. As we reach low redshifts, where dark energy dominates the
universal expansion, $\alpha _{b}$ becomes a constant \citep{bsbm,mota1},
but $\Delta _{c}$ continues to increase, as will $\alpha _{c}$ \citep{mota3}%
. The growth of $\alpha _{v}$ becomes steeper as we go from a dark-energy
fluid with $w_{\phi }=-0.6$ to the $\Lambda $-like case of $w_{\phi }=-1$.
The intermediate situation is where $w_{\phi }=-0.8$, due to the dependence
of $\Delta _{c}$ on $w_{\phi }$.

\begin{figure}%[tbp]
\epsfig{file=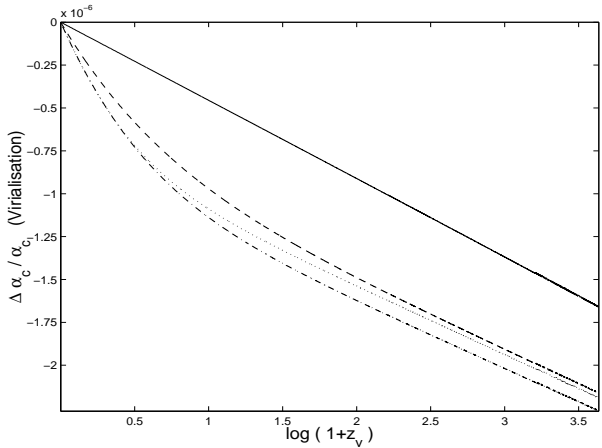,height=6cm,width=8cm}
\caption{\textit{Variation of $\Delta \protect\alpha /\protect\alpha $
inside clusters as a function of the epoch of virialisation, $\log (1+z_{v})$%
. The solid line corresponds to the standard ($\Lambda =0$) $CDM$, the
dashed line is the $\Lambda CDM$, the dashed-dotted line corresponds to a $%
wCDM$ model with $w_{\protect\phi }=-0.8$, the dotted line corresponds to a $%
wCDM$ model with $w_{\protect\phi }=-0.6$. In all models, the initial
conditions were set in order to have $\protect\alpha _{c}(z=0)=\protect%
\alpha _{0}$ and to satisfy $\Delta \protect\alpha /\protect\alpha \approx
-5.4\times 10^{-6}$ at $3.5\geq |z-z_{v}|\geq 0.5$.}}
\label{modelsdeltatimevirial}
\end{figure}

\begin{figure}%[tbp]
\epsfig{file=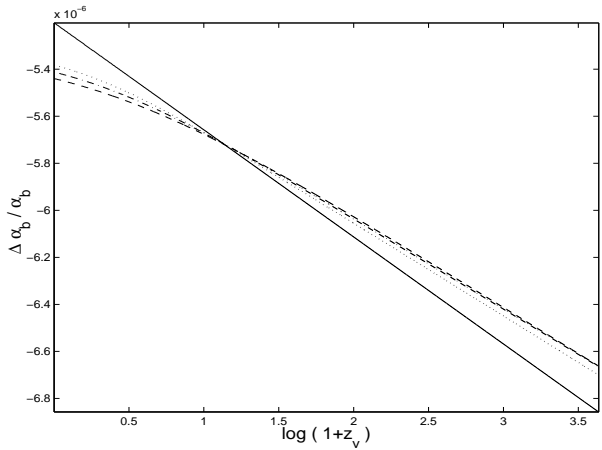,height=6cm,width=8cm}
\caption{\textit{Variation of $\Delta \protect\alpha /\protect\alpha $ in
the background at the epoch of cluster virialisation, $\log (1+z_{v})$. The
solid line corresponds to the standard ($\Lambda =0$) $CDM$, the dashed line
is the $\Lambda CDM$ model, the dashed-dotted line is $w_{\protect\phi %
}=-0.8 $, the dotted line ($w_{\protect\phi }=-0.6$). In all models, the
initial conditions were set in order to have $\protect\alpha _{c}(z=0)=%
\protect\alpha _{0}$ and to satisfy $\Delta \protect\alpha /\protect\alpha %
\approx -5.4\times 10^{-6}$ at $3.5\geq |z-z_{v}|\geq 0.5$.}}
\label{modelsdeltatimevirialback}
\end{figure}

Similar conclusions can be drawn with respect to the 'time shift' of the
fine structure 'constant', ($\Delta \alpha /\alpha $), at virialisation, see
Figures \ref{modelsdeltatimevirial} and \ref{modelsdeltatimevirialback}. 

\subsection{Spatial variations in $\protect\alpha$}

\begin{figure}%[tbp]
\epsfig{file=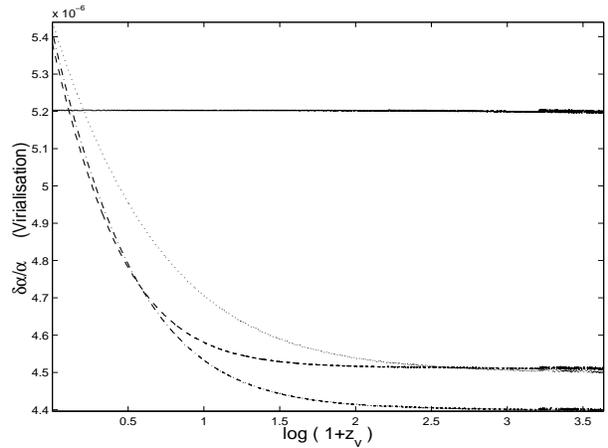,height=6cm,width=8cm}
\caption{\textit{Variation of $\protect\delta \protect\alpha /\protect\alpha 
$ as a function of $\log (1+z_{v})$. Solid line corresponds to the standard (%
$\Lambda =0$) $CDM$, dashed line is the $\Lambda CDM$, dashed-dotted line
corresponds to a $wCDM$ model with $w_{\protect\phi }=-0.8$, dotted line
corresponds to a $wCDM$ model with $w_{\protect\phi }=-0.6$. In all the
models, the initial conditions were set in order to have $\protect\alpha %
_{c}(z=0)=\protect\alpha _{0}$ and to satisfy $\Delta \protect\alpha /%
\protect\alpha \approx -5.4\times 10^{-6}$ at $3.5\geq |z-z_{v}|\geq 0.5$.}}
\label{modelsdeltaa}
\end{figure}

Spatial variations in $\alpha $ will be dramatically different when
comparing the standard $CDM$ model with the $wCDM$ models. In the $SCDM$
model, the difference between the fine structure 'constant' in a virialised
cluster ($\alpha _{v}$) and in the background ($\alpha _{b}$) will always be
the same, $\delta \alpha /\alpha \approx 5.2\times 10^{-6}$, independently
of the redshift at which we measure it, Figure \ref{modelsdeltaa}. 
Again, this is because, in a $SCDM$ model, $\Delta _{c}$ is always a
constant independent of the redshift at which virialisation occurs. The
constancy of $\delta \alpha /\alpha $ is a signature of the $SCDM$
structure-formation model, and it may even provide a means to rule out the $%
SCDM$ model completely if, when comparing the value of $\delta \alpha
/\alpha $ in two different clusters, we do not find the same value,
independently of $z_{v}$. %Of course, there are better ways to do that,
%by using the CMBR power spectrum  \citep{wmap} or
%clusters abundances  \citep{lahav,lahav1}. 
A similar result is found for the case of a dark-energy structure formation
model at high redshifts where $\delta \alpha /\alpha $ will be constant.
This behaviour is expected since at high redshift any $wCDM$ model is
asymptotically equivalent to standard $CDM$.

\begin{figure}%[tbp]
\epsfig{file=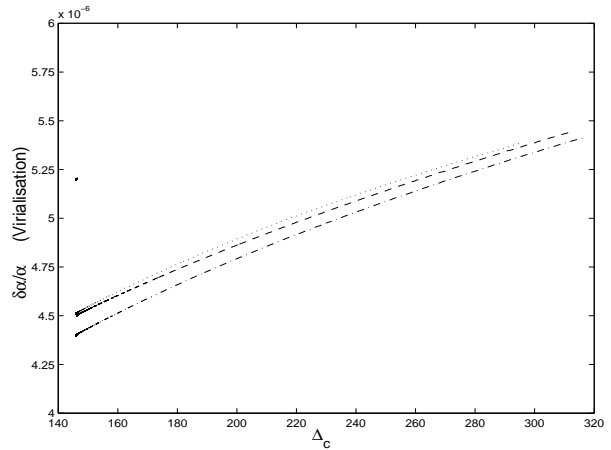,height=6cm,width=8cm}
\caption{\textit{Variation of $\protect\delta \protect\alpha /\protect\alpha 
$ as a function as a function of $\Delta _{c}$. The single-point corresponds
to the standard ($\Lambda =0$) $CDM$, the dashed line is the $\Lambda CDM$,
the dash-dot line corresponds to a $wCDM$ model with $w_{\protect\phi }=-0.8$%
, dotted-line corresponds to a $wCDM$ model with $w_{\protect\phi }=-0.6$.
In all the models, the initial conditions were set in order to have $\protect%
\alpha _{c}(z=0)=\protect\alpha _{0}$ and to satisfy $\Delta \protect\alpha /%
\protect\alpha \approx -5.4\times 10^{-6}$ at $3.5\geq |z-z_{v}|\geq 0.5$.}}
\label{modelsdeltarho}
\end{figure}

\begin{figure}%[tbp]
\epsfig{file=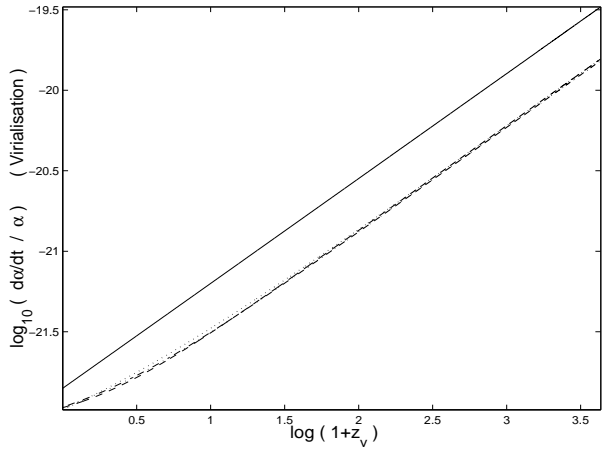,height=6cm,width=8cm}
\caption{\textit{Variation of $log_{10}(\dot{\protect\alpha}/\protect\alpha %
) $ as a function of $log(1+z_{v})$. Solid line corresponds to the Standard (%
$\Lambda =0$) $CDM$, dashed-line is the $\Lambda CDM$, dashed-dotted
corresponds to a $wCDM$ model with $w_{\protect\phi }=-0.8$, dotted line
corresponds to a $wCDM$ model with $w_{\protect\phi }=-0.6$. In all the
models, the initial conditions were set in order to have $\protect\alpha %
_{c}(z=0)=\protect\alpha _{0}$ and to satisfy $\Delta \protect\alpha /%
\protect\alpha \approx -5.4\times 10^{-6}$ at $3.5\geq |z-z_{v}|\geq 0.5$.}}
\label{modelspsidot}
\end{figure}

As expected, it is at low redshifts that the difference between $wCDM$ and $%
SCDM$ emerges. When comparing virialised regions at low redshifts, $\delta
\alpha /\alpha $ will increase in a $wCDM$ model as we approach $z=0$. This
is due to an increase of the density contrast of the virialised regions, $%
\Delta _{c}$, and the approach to a constant value of $\alpha _{b}$, see
Figure \ref{modelsdeltarho}. 
In general, the growth will be steeper for smaller values of $w_{\phi }$,
although there will be parameter degeneracies between the behaviour of
different models, which ensure that there is no simple relation between $%
\delta \alpha /\alpha $ and the dark-energy equation of state. Note that,
independently of the structure formation model we use, $\dot{\alpha}%
_{c}/\alpha _{c}$ at virialisation is always a decreasing function of time,
as shown in Figure \ref{modelspsidot}. 
%
%This result just confirms our claim that, if variations
%in $\alpha $ are so small for such a wide range of redshifts, we can assume
%that the difference between the value of $\alpha _{c}$ at $z_{v}$ and at $z=0
%$ will be negligible. Therefore it is a good approximation to assume that
%the time-evolution of both $\alpha _{c}$ and of the cluster will cease after
%virialisation.

\section{The Dependence on the Coupling of $\protect\alpha $-Variation to
the Matter Fields}

The evolution of $\alpha $ in the background and inside clusters depends
mainly on the dominant equation of state of the universe and the \textit{sign%
} of the coupling constant $\zeta /\omega ,$ which is determined by the
theory and the dark matter's identity. %
%Here, we have assumed
%that $\omega =1$, so all the dependence is in $\zeta $. 
%
As was shown in \citep{bsbm,mota1}, $\alpha _{b}$ will be nearly constant
for an accelerated expansion and also during the radiation era far from the
initial singularity (where the kinetic term, $\rho _{\psi }$, can dominate). 
\emph{\ }Slow evolution of $\alpha $ will occur during the dust-dominated
epoch, where $\alpha $ increases logarithmically in time for $\zeta <0$.
When $\zeta $ is negative, $\alpha $ will be a slowly growing function of
time but $\alpha $ will fall rapidly (even during a curvature-dominated era)
for $\zeta $ positive \citep{bsbm}. A similar behaviour is found for the
evolution of the fine structure 'constant' inside overdensities. Thus, we
see that a slow change in $\alpha $, cut off by the accelerated expansion at
low redshift, that may be required by the data, demands that $\zeta <0$ in
the cosmological background.

The sign of $\zeta $ is determined by the physical character of the matter
that carries electromagnetic coupling. If it is dominated by magnetic energy
then $\zeta <0$, if not then $\zeta >0.$ Baryons will usually have a
positive $\zeta $ (although Bekenstein has argued for negative baryonic $%
\zeta $ in ref. \citep{bk}, but see \citep{dam}), in particular $\zeta
\approx 10^{-4}$ for neutrons and protons. Dark matter may have negative
values of $\zeta $, for instance superconducting cosmic strings have$\ \zeta
\approx -1$.

In the previous section, we have chosen the sign of $\zeta $ to be negative
so $\alpha $ is a slowly-growing function in time during the era of dust
domination. This was done in order to match the latest observations which
suggest that $\alpha $ had a smaller value in the past %
\citep{murphy,murphy1,murphy2,murphy3}. This is a good approximation, since
we have been studying the cosmological evolution of $\alpha $ during
large-scale structure formation, when dark matter dominates. However, we
know that on sufficiently small scales the dark matter will become dominated
by a baryonic contribution for which $\zeta >0.$ The transition in the
dominant form of total density, from non-baryonic to baryonic as one goes
from large to small scales requires a significant evolution in the magnitude
and sign of $\zeta /\omega $. This inhomogeneity will create distinctive
behaviours in the evolution of the fine structure 'constant' and will be
studied in more elsewhere. It is clear that a change in the sign of $\zeta
/\omega $ will lead to a completely different type of evolution for $\alpha $%
, although the expected variations in the sign of $\zeta /\omega $ will
occur on scales much smaller than those to which we are applying the
spherical collapse model here. Hence, we will only investigate the effects
of changing the absolute value of the coupling, $|\zeta /\omega |$, for the
evolution of the fine structure 'constant'.

%In order to study the dependence on the coupling $\zeta/\omega$ 
%numerical simulations were
%performed using a $\Lambda CDM$ model and for the following values of the coupling to the matter
%fields,   
%$\zeta/\omega=-10^{-4}, -4 \times 10^{-4}, -7 \times10^{-4}$ and
%$-10^{-5}$. We have set the initial conditions to give $\alpha_c
%(z=0)=\alpha_0$ and to satisfy  $\Delta \alpha /\alpha \approx -5.4\times 
%10^{-6}$ at $3\geq |z-z_{v}|\geq 1$, in the case where
%$\zeta/\omega=-7\times10^{-4}$.  

\begin{figure}%[tbp]
\epsfig{file=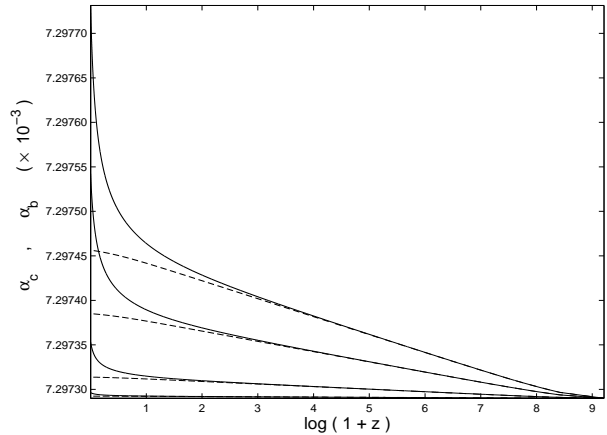,height=6cm,width=8cm}
\caption{\textit{Evolution of $\protect\alpha $ in the background
(dashed-line) and inside a cluster (Solid line) as a function of $\log (1+z)$%
. For $\protect\zeta /\protect\omega =-10^{-8},-4\times 10^{-8},-7\times
10^{-8}$ and $-10^{-9}$. The lower curves correspond to a lower value of $%
\protect\zeta /\protect\omega $. The initial conditions were set in order to
have $\protect\alpha _{c}(z=0)=\protect\alpha _{0}$ and to satisfy $\Delta 
\protect\alpha /\protect\alpha =-5.4\times 10^{-6}$ at $3\geq |z-z_{v}|\geq
1 $, in the case where $\protect\zeta =-7\times 10^{-8}$. }}
\label{zetasalphacb}
\end{figure}

\begin{figure}%[tbp]
\epsfig{file=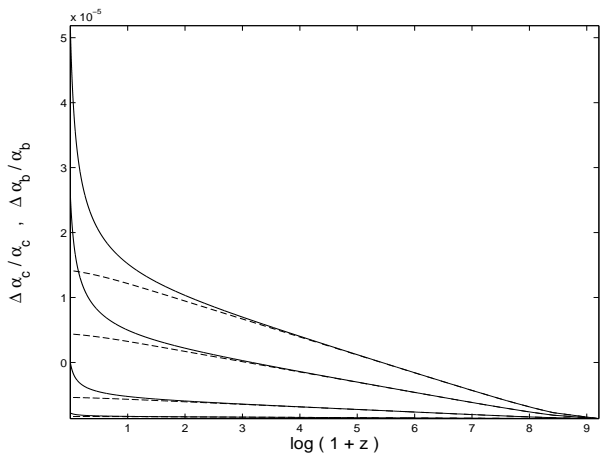,height=6cm,width=8cm}
\caption{\textit{Evolution of $\Delta \protect\alpha /\protect\alpha $ in
the background (dashed-line) and inside a cluster (Solid line) as a function
of $\log (1+z)$. For $\protect\zeta /\protect\omega =-10^{-8},-4\times
10^{-8},-7\times 10^{-8}$ and $-10^{-9}$. The lower curves correspond to a
lower value of $\protect\zeta /\protect\omega $. The initial conditions were
set in order to have $\protect\alpha _{c}(z=0)=\protect\alpha _{0}$ and to
satisfy $\Delta \protect\alpha /\protect\alpha =-5.4\times 10^{-6}$ at $%
3\geq |z-z_{v}|\geq 1$, in the case where $\protect\zeta =-7\times 10^{-8}$. 
}}
\label{zetasalphacbback}
\end{figure}

From Figure \ref{zetasalphacb} it is clear that the rate of changes in $%
\alpha _{b}$ and $\alpha _{c}$ will be functions of the absolute value of $%
\zeta /\omega $. Smaller values of $|\zeta /\omega |$ lead to a slower
variation of $\alpha $. A similar behaviour is found for the time variations
in $\Delta \alpha /\alpha $, see Figure \ref{zetasalphacbback}. 

\begin{figure}%[tbp]
\epsfig{file=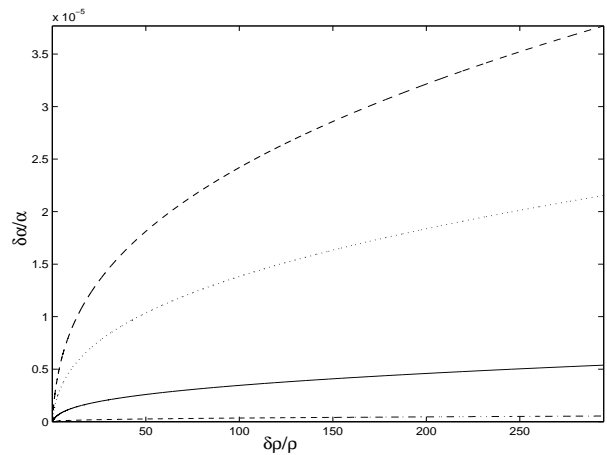,height=6cm,width=8cm}
\caption{\textit{Evolution of $\protect\delta \protect\alpha /\protect\alpha 
$ as a function $\protect\delta \protect\rho /\protect\rho $. For $\protect%
\zeta /\protect\omega =-10^{-8},-4\times 10^{-8},-7\times 10^{-8}$ and $%
-10^{-9}$. The lower curves correspond to a lower value of $\protect\zeta /%
\protect\omega $. The initial conditions were set in order to have $\protect%
\alpha _{c}(z=0)=\protect\alpha _{0}$ and to satisfy $\Delta \protect\alpha /%
\protect\alpha \approx -5.4\times 10^{-6}$ at $3\geq |z-z_{v}|\geq 1$, in
the case where $\protect\zeta =-7\times 10^{-8}$. }}
\label{zetasdeltaa}
\end{figure}

The faster variation in $\alpha $ and $\Delta \alpha /\alpha $ for higher
values of $|\zeta /\omega |$ is also a common feature for $\delta \alpha
/\alpha $, see Figure \ref{zetasdeltaa}. This is expected. A stronger
coupling to the matter fields would naturally lead to a stronger dependence
on the matter inhomogeneities, and in particular on their density contrast, $%
\Delta _{c}$. %Notice however
%that 
%this difference is almost negligible during most of the history of the
%cluster formation, figure \ref{zetasdeltarho}. It becomes important only
%during the last stages of its evolution, near the turnaround time  \citep{mota3}.
%
%
%\begin{figure}[htbp!]
%\epsfig{file=zetasdeltaa.eps,height=6yyyyyyyyyyyyyycm,width=8cm}
%\caption{{\protect {\it Evolution of
%$\protect\delta\alpha/\alpha$  inside a cluster as a function 
%of $\log (a)$.  For  $\zeta/\omega=-10^{-4}, -4 \times 10^{-4}, -7 \times10^{-4}$ and
%$-10^{-5}$. The lower curves correspond to a lower value of
%$\zeta/\omega$. The initial condition was set in order to have  $\alpha_c
%(z=0)=\alpha_0$ and to satisfy  $\Delta \alpha /\alpha \approx -5.4\times 
%10^{-6}$ at $3\geq |z-z_{v}|\geq 1$, in the case where
%$\zeta=-7\times10^{-4}$. }}}
%\label{zetasdeltarho}
%\end{figure}

In reality, the dependence of $\alpha $ on the coupling $\zeta /\omega $ has
a degeneracy with respect to the initial condition chosen for $\psi $. This
is clear from the scale invariance of equation (\ref{psidot}) under linear
shifts in the value of $\psi \rightarrow \psi +const$ and rescaling of $%
\zeta /\omega $ and $t$. It is always possible, to obtain the same evolution
(and rate of change) of $\alpha $ and $\Delta \alpha /\alpha $ for any other
value of $|\zeta /\omega |$; see for example Tables \ref{zeta1},\ref{zeta4},%
\ref{zeta6} and \ref{zeta10}, where we have tabulated the shifts $\Delta
\alpha /\alpha ,\delta \alpha /\alpha ,$ and time change, $\dot{\alpha}%
/\alpha ,$ obtained in the clusters and in the background for various
numerical choices of $\zeta /\omega $ and virialisation redshift $z_{v}$.
The observed Oklo and $\beta $-decay constraints on variations in $\alpha $
are highlighted in italic and boldface, respectively. 
\begin{table*}%[tbp]
\begin{minipage}{150mm}
\begin{center}
%{\large{$\frac{\zeta}{\omega}=-10^{-1}$}}
\begin{tabular}{|c|c|c|c|c|c|c|}
\hline
$z_v$ & $\frac{\Delta\alpha}{\alpha} \vert_{b}\times 10^{-6}$ & $\frac{%
\Delta\alpha}{\alpha} \vert_{c}\times 10^{-7}$ & $\frac{\dot\alpha}{\alpha}
\vert_{b}\times 10^{-24}$ & $\frac{\dot\alpha}{\alpha} \vert_{c}\times
10^{-22}$ & $\frac{\delta\alpha}{\alpha}\times 10^{-6}$ \\ \hline
$0.00$ & $-5.381$ & $0.000$ & $0.361$ & $1.068$ & $5.38$ \\ 
$\mathit{0.09}$ & $\mathit{-5.397}$ & $\mathit{\ -1.618}$ & $\mathit{\ 0.427}
$ & $\mathit{\ 1.128}$ & $\mathit{\ 5.24}$ \\ 
$0.18$ & $-5.412$ & $-2.889$ & $0.493$ & $1.192$ & $5.12$ \\ 
%$0.26$ & $-5.427$ & $-3.915$ & $0.558$ & $1.260$ & $5.04$ \\ 
$0.34$ & $-5.442$ & $-4.769$ & $0.623$ & $1.329$ & $4.97$ \\ 
$\mathbf{0.41}$ & $\mathbf{-5.456}$ & $\mathbf{-5.485}$ & $\mathbf{\ 0.688}$
& $\mathbf{\ 1.401}$ & $\mathbf{\ 4.91}$ \\ 
%$0.49$ & $-5.470$ & $-6.097$ & $0.753$ & $1.475$ & $4.86$ \\ 
$0.51$ & $-5.474$ & $-6.271$ & $0.773$ & $1.498$ & $4.85$ \\ 
$1.07$ & $-5.572$ & $-9.139$ & $1.303$ & $2.161$ & $4.66$ \\ 
%$1.51$ & $-5.639$ & $-10.428$ & $1.773$ & $2.793$ & $4.60$ \\ 
$2.14$ & $-5.722$ & $-11.658$ & $2.508$ & $3.821$ & $4.56$ \\ 
%$2.54$ & $-5.767$ & $-12.237$ & $3.009$ & $4.535$ & $4.54$ \\ 
$3.31$ & $-5.843$ & $-13.129$ & $4.055$ & $6.043$ & $4.53$ \\ 
%$4.07$ & $-5.906$ & $-13.835$ & $5.190$ & $7.690$ & $4.52$ \\ 
$4.83$ & $-5.961$ & $-14.415$ & $6.407$ & $9.466$ & $4.52$ \\ \hline
\end{tabular}%
\end{center}
\caption{Time and space
variations in $\protect\alpha $ obtained for the corresponding redshifts of
virialisation, $z_{v}$, for $\protect\zeta /\protect\omega =-10^{-1}$. We
have assumed a $\Lambda CDM$ model. The indexes $^{\prime }b^{\prime }$ and $%
^{\prime }c^{\prime }$, stand for background and cluster respectively. The
italic and bold entries correspond approximately to the level of the Oklo
and $\protect\beta $-decay rate constraints, respectively. The quasar
absorption spectra observations correspond to the values of $\Delta \protect%
\alpha /\protect\alpha |_{b}$. The initial conditions were set in order to
have $\protect\alpha _{c}(z=0)=\protect\alpha _{0}$.}
\label{zeta1}
\end{minipage}
\end{table*}

\begin{table*}%[tbp]
\begin{minipage}{150mm}
\begin{center}
%{\large{$\frac{\zeta}{\omega}=-10^{-4}$}}
\begin{tabular}{|c|c|c|c|c|c|c|}
\hline
$z_v$ & $\frac{\Delta\alpha}{\alpha} \vert_{b}\times 10^{-6}$ & $\frac{%
\Delta\alpha}{\alpha} \vert_{c}\times 10^{-7}$ & $\frac{\dot\alpha}{\alpha}
\vert_{b}\times 10^{-24}$ & $\frac{\dot\alpha}{\alpha} \vert_{c}\times
10^{-22}$ & $\frac{\delta\alpha}{\alpha}\times 10^{-6}$ \\ \hline
$0.00$ & $-5.382$ & $0.000$ & $0.361$ & $1.068$ & $5.38$ \\ 
$\mathit{0.09}$ & $\mathit{-5.398}$ & $\mathit{\ -1.618}$ & $\mathit{\ 0.427}
$ & $\mathit{\ 1.128}$ & $\mathit{\ 5.24}$ \\ 
$0.18$ & $-5.414$ & $-2.889$ & $0.493$ & $1.193$ & $5.12$ \\ 
%$0.26$ & $-5.429$ & $-3.916$ & $0.558$ & $1.260$ & $5.04$ \\ 
$0.34$ & $-5.443$ & $-4.769$ & $0.623$ & $1.329$ & $4.97$ \\ 
$\mathbf{0.41}$ & $\mathbf{-5.458}$ & $\mathbf{\ -5.486}$ & $\mathbf{\ 0.688}
$ & $\mathbf{\ 1.401}$ & $\mathbf{\ 4.91}$ \\ 
%$0.49$ & $-5.471$ & $-6.098$ & $0.753$ & $1.475$ & $4.86$ \\ 
$0.51$ & $-5.476$ & $-6.272$ & $0.773$ & $1.498$ & $4.85$ \\ 
$1.07$ & $-5.573$ & $-9.141$ & $1.303$ & $2.161$ & $4.66$ \\ 
%$1.51$ & $-5.641$ & $-10.430$ & $1.773$ & $2.794$ & $4.60$ \\ 
$2.14$ & $-5.723$ & $-11.660$ & $2.508$ & $3.822$ & $4.56$ \\ 
%$2.54$ & $-5.769$ & $-12.239$ & $3.009$ & $4.536$ & $4.54$ \\ 
$3.31$ & $-5.844$ & $-13.132$ & $4.056$ & $6.045$ & $4.53$ \\ 
%$4.07$ & $-5.908$ & $-13.837$ & $5.191$ & $7.692$ & $4.52$ \\ 
$4.83$ & $-5.962$ & $-14.418$ & $6.408$ & $9.469$ & $4.52$ \\ \hline
\end{tabular}%
\end{center}
\caption{Time and space
variations in $\protect\alpha $ for the corresponding redshifts of
virialisation, $z_{v},$ with $\protect\zeta /\protect\omega =-10^{-5}$ .We
assumed a $\Lambda CDM$ model. The indices $^{\prime }b^{\prime }$ and $%
^{\prime }c^{\prime }$, stand for background and cluster respectively. The
italic and bold entries correspond approximately to the Oklo and $\protect%
\beta $-decay rate constraints, respectively. The initial conditions were
set in order to have $\protect\alpha _{c}(z=0)=\protect\alpha _{0}$.}
\label{zeta4}
\end{minipage}
\end{table*}

\begin{table*}%[tbp]
\begin{minipage}{150mm}
\begin{center}
%{\large{$\frac{\zeta}{\omega}=-10^{-6}$}}
\begin{tabular}{|c|c|c|c|c|c|c|}
\hline
$z_v$ & $\frac{\Delta\alpha}{\alpha} \vert_{b}\times 10^{-6}$ & $\frac{%
\Delta\alpha}{\alpha} \vert_{c}\times 10^{-7}$ & $\frac{\dot\alpha}{\alpha}
\vert_{b}\times 10^{-24}$ & $\frac{\dot\alpha}{\alpha} \vert_{c}\times
10^{-22}$ & $\frac{\delta\alpha}{\alpha}\times 10^{-6}$ \\ \hline
$0.00$ & $-5.381$ & $0.000$ & $0.361$ & $1.068$ & $5.38$ \\ 
$\mathit{0.09}$ & $\mathit{-5.397}$ & $\mathit{\ -1.617}$ & $\mathit{\ 0.427}
$ & $\mathit{\ 1.128}$ & $\mathit{\ 5.24}$ \\ 
$0.18$ & $-5.413$ & $-2.889$ & $0.493$ & $1.192$ & $5.12$ \\ 
%$0.26$ & $-5.428$ & $-3.915$ & $0.558$ & $1.260$ & $5.04$ \\ 
$0.34$ & $-5.442$ & $-4.768$ & $0.623$ & $1.329$ & $4.97$ \\ 
$\mathbf{0.41}$ & $\mathbf{-5.457}$ & $\mathbf{\ -5.485}$ & $\mathbf{\ 0.688}
$ & $\mathbf{\ 1.401}$ & $\mathbf{\ 4.91}$ \\ 
%$0.49$ & $-5.470$ & $-6.097$ & $0.753$ & $1.475$ & $4.86$ \\ 
$0.51$ & $-5.475$ & $-6.271$ & $0.773$ & $1.498$ & $4.85$ \\ 
$1.07$ & $-5.572$ & $-9.139$ & $1.303$ & $2.161$ & $4.66$ \\ 
%$1.51$ & $-5.640$ & $-10.428$ & $1.773$ & $2.794$ & $4.60$ \\ 
$2.14$ & $-5.722$ & $-11.658$ & $2.508$ & $3.822$ & $4.56$ \\ 
%$2.54$ & $-5.768$ & $-12.237$ & $3.009$ & $4.535$ & $4.54$ \\ 
$3.31$ & $-5.843$ & $-13.129$ & $4.056$ & $6.044$ & $4.53$ \\ 
%$4.07$ & $-5.907$ & $-13.835$ & $5.190$ & $7.690$ & $4.52$ \\ 
$4.83$ & $-5.961$ & $-14.415$ & $6.407$ & $9.467$ & $4.52$ \\ \hline
\end{tabular}%
\end{center}
\caption{Time and space
variations in $\protect\alpha $ for the corresponding redshifts of
virialisation, $z_{v},$ for $\protect\zeta /\protect\omega =-10^{-9}$ .We
have assumed a $\Lambda CDM$ model. The indices $^{\prime }b^{\prime }$ and $%
^{\prime }c^{\prime }$, stand for background and cluster respectively. The
italic and bold entries correspond approximately to the Oklo and to the $%
\protect\beta $-decay rate constraints, respectively. The initial conditions
were set in order to have $\protect\alpha _{c}(z=0)=\protect\alpha _{0}$.}
\label{zeta6}
\end{minipage}
\end{table*}

\begin{table*}%[tbp]
\begin{minipage}{150mm}
\begin{center}
%{\large{$\frac{\zeta}{\omega}=-10^{-10}$}}
\begin{tabular}{|c|c|c|c|c|c|c|}
\hline
$z_v$ & $\frac{\Delta\alpha}{\alpha} \vert_{b}\times 10^{-6}$ & $\frac{%
\Delta\alpha}{\alpha} \vert_{c}\times 10^{-7}$ & $\frac{\dot\alpha}{\alpha}
\vert_{b}\times 10^{-24}$ & $\frac{\dot\alpha}{\alpha} \vert_{c}\times
10^{-22}$ & $\frac{\delta\alpha}{\alpha}\times 10^{-6}$ \\ \hline
$0.00$ & $-5.385$ & $0.000$ & $0.361$ & $1.068$ & $5.38$ \\ 
$\mathit{0.09}$ & $\mathit{-5.401}$ & $\mathit{\ -1.619}$ & $\mathit{\ 0.428}
$ & $\mathit{\ 1.129}$ & $\mathit{\ 5.24}$ \\ 
$0.18$ & $-5.416$ & $-2.891$ & $0.493$ & $1.193$ & $5.13$ \\ 
%$0.26$ & $-5.431$ & $-3.918$ & $0.559$ & $1.261$ & $5.04$ \\ 
$0.34$ & $-5.446$ & $-4.772$ & $0.624$ & $1.330$ & $4.97$ \\ 
$\mathbf{0.41}$ & $\mathbf{-5.460}$ & $\mathbf{\ -5.489}$ & $\mathbf{\ 0.689}
$ & $\mathbf{\ 1.402}$ & $\mathbf{\ 4.91}$ \\ 
%$0.49$ & $-5.474$ & $-6.101$ & $0.753$ & $1.476$ & $4.86$ \\ 
$0.51$ & $-5.478$ & $-6.275$ & $0.773$ & $1.499$ & $4.85$ \\ 
$1.07$ & $-5.576$ & $-9.145$ & $1.304$ & $2.162$ & $4.66$ \\ 
%$1.51$ & $-5.644$ & $-10.435$ & $1.774$ & $2.795$ & $4.60$ \\ 
$2.14$ & $-5.726$ & $-11.666$ & $2.510$ & $3.824$ & $4.56$ \\ 
%$2.54$ & $-5.771$ & $-12.245$ & $3.011$ & $4.538$ & $4.55$ \\ 
$3.31$ & $-5.847$ & $-13.138$ & $4.058$ & $6.048$ & $4.53$ \\ 
%$4.07$ & $-5.911$ & $-13.844$ & $5.193$ & $7.695$ & $4.53$ \\ 
$4.83$ & $-5.965$ & $-14.425$ & $6.412$ & $9.473$ & $4.52$ \\ \hline
\end{tabular}%
\end{center}
\caption{Time and space
variations in $\protect\alpha $ for the redshifts of virialisation, $z_{v}$,
for $\protect\zeta /\protect\omega =-10^{-13}$. We have assumed a $\Lambda
CDM$ model. The indices $^{\prime }b^{\prime }$ and $^{\prime }c^{\prime }$,
stand for background and cluster respectively. The italic and bold entries
correspond approximately to the Oklo and $\protect\beta $-decay rate
constraints, respectively. The initial conditions were set in order to have $%
\protect\alpha _{c}(z=0)=\protect\alpha _{0}$.}
\label{zeta10}
\end{minipage}
\end{table*}

In the plots of this section, we have chosen the value $\zeta /\omega
=-7\times 10^{-8}$ to be the one which satisfies the current observations,
but we could have used any value of $|\zeta /\omega |$ because of the
invariance under rescalings. However, once we set the initial condition for
a given value of $\zeta /\omega $, any deviation from that value leads to
quite different future variations in $\alpha $. This can occur if there are
regions of the universe where the dominant matter has a different nature,
and and do possesses a different value (and even sign) of $\zeta /\omega $
to that in our solar system. The evolution of $\alpha $ in those regions may
then be different from the one that led to the value of $\alpha
_{c}(z=z_{v})=\alpha _{0}$ on Earth. %This dependence may give  important
%constraints on the nature of dark matter, depending
%on the varying-$\alpha$ model and the matter fields to which $\alpha$ is coupled.

It is important to note that, due to the degeneracy between the initial
condition and the coupling to the matter fields, there is no way to avoid
evolving differences in $\alpha $ variations between the overdensities and
the background. The difference will be of the same order of magnitude as the
effects indicated by the recent quasar absorption-line data. This is not a
coincidence and it is related to the fact that we have normalised $\alpha
(z=z_{v})=\alpha _{0}$ on Earth and $\Delta \alpha /\alpha =5.4\times
10^{-6} $ at $3.5\geq |z-z_{v}|\geq 0.5$. So, any varying-$\alpha $ model
that uses these normalisations will create a similar difference in $\alpha $
evolution between the overdensities and the background, independently of the
coupling $\zeta /\omega $.

We might ask: how model-independent are these results? In this connection it
is interesting to note that even with a zero coupling to the matter fields
(which is unrealistic), $\zeta =0$, there is no way to avoid difference
arising between the evolution of $\alpha $ in the background and in the
cluster overdensities. While the background\ expansion has a
monotonically-increasing scale factor, $a(t)$, the overdensities will have a
scale radius, $R(t)$, which will eventually collapse at a finite time. For
instance, in the case where $\zeta =0$, equations (\ref{psidot}) and (\ref%
{psidotcluster1}) can be automatically integrated to give: 
\begin{equation}
\dot{\psi} \propto a^{-3} \qquad , \qquad \dot{\psi}_{c} \propto R^{-3}
\end{equation}
The difference between those two solutions clearly increases, especially
after the turnaround of the overdensity, when that region starts to
collapse. As the collapse proceeds, the bigger will be the difference
between the background and the overdensity. Variations in $\alpha $ between
the background and the overdensities are therefore\ quite natural although
they have always been ignored in studies of varying constants in cosmology. 
%Note, that even in the background, $\alpha$ and its time
%variation, $\Delta\alpha/\alpha$, will
%present different behaviours, depending on the  equation of state
%of the universe at that epoch. For instance, during a radiation
%dominated epoch $\alpha$ will be a constant and so
%$\Delta\alpha/\alpha=0$, while during a dust 
%dominated one $\alpha$ will be a slowly growing function of time and
%so $\Delta\alpha/\alpha\neq0$ ($\approx-10^{-6}$ for $z=0.5-3.5$). 

\section{Discussion of the Results and Observational Constraints}

The development of matter inhomogeneities in our universe affects the
cosmological evolution of the fine structure 'constant' \citep{mota2,mota3}.
Therefore, variations in $\alpha $ depend on nature of its coupling to the
matter fields and the detailed large-scale structure formation model.
Large-scale structure formation models depend in turn on the dark-energy
equation of state. This dependence is particularly strong at low redshifts,
when dark energy dominates the density of the universe \citep{lahav,lahav1}.
Using the BSBM varying-$\alpha $ theory, and the simplest spherical collapse
model, we have studied the effects of the dark-energy equation of state and
the coupling to the matter fields on the evolution of the fine structure
'constant'. We have compared the evolution of $\alpha $ inside virialised
overdensities, using the standard ($\Lambda =0$) $CDM$ model of structure
formation and dark-energy modification ($wCDM$). It was shown that,
independently of the model of structure formation one considers, there is
always a spatial contrast, $\delta \alpha /\alpha $, between $\alpha $ in an
overdensity and in the background. In a $SCDM$ model, $\delta \alpha /\alpha 
$ is always a constant, independent of the virialisation redshift, see Table %
\ref{darkscdm}. In the case of a $wCDM$ model, especially at low redshifts,
the spatial contrast depends on the time when virialisation occurs and the
equation of state of the dark energy. At high redshifts, when the $wCDM$
model becomes asymptotically equivalent to the $SCDM$ one, $\delta \alpha
/\alpha $ is a constant. At low redshifts, when dark energy starts to
dominate, the difference between $\alpha $ in a cluster and in the
background grows. The growth rate is proportional to $|w_{\phi }|$, see
Tables \ref{darkw6cdm}, \ref{darkw8cdm} and \ref{darklcdm}. These
differences in the behaviour of the fine structure 'constant', its 'time
shift density contrast' ($\Delta \alpha /\alpha $) and its 'spatial density
contrast' ($\delta \alpha /\alpha $) could help us to distinguish among
different dark-energy models of structure formation at low redshifts.

\begin{table*}%[tbp]
\begin{minipage}{150mm}
%\begin{ruledtabular}
\par
\begin{center}
%{\large{$(\Lambda=0)$ CDM }}
\begin{tabular}{|c|c|c|c|c|c|c|}
\hline
$z_v$ & $\frac{\Delta\alpha}{\alpha} \vert_{b}\times 10^{-6}$ & $\frac{%
\Delta\alpha}{\alpha} \vert_{c}\times 10^{-7}$ & $\frac{\dot\alpha}{\alpha}
\vert_{b}\times 10^{-24}$ & $\frac{\dot\alpha}{\alpha} \vert_{c}\times
10^{-22}$ & $\frac{\delta\alpha}{\alpha}\times 10^{-6}$ \\ \hline
$0.00$ & $-5.203$ & $0.000$ & $0.959$ & $1.409$ & $5.20$ \\ 
$0.06$ & $-5.229$ & $-0.266$ & $1.047$ & $1.537$ & $5.20$ \\ 
$\mathit{0.12}$ & $\mathit{-5.254}$ & $\mathit{-0.517}$ & $\mathit{\ 1.137}$
& $\mathit{\ 1.670}$ & $\mathit{\ 5.20}$ \\ 
%$0.30$ & $-5.322$ & $-1.196$ & $1.422$ & $2.088$ & $5.20$ \\ 
$0.38$ & $-5.349$ & $-1.463$ & $1.553$ & $2.281$ & $5.20$ \\ 
$\mathbf{0.46}$ & $\mathbf{-5.374}$ & $\mathbf{-1.717}$ & $\mathbf{\ 1.690}$
& $\mathbf{\ 2.481}$ & $\mathbf{\ 5.20}$ \\ 
$0.54$ & $-5.399$ & $-1.962$ & $1.830$ & $2.687$ & $5.20$ \\ 
$1.07$ & $-5.534$ & $-3.313$ & $2.859$ & $4.198$ & $5.20$ \\ 
%$1.52$ & $-5.623$ & $-4.200$ & $3.827$ & $5.620$ & $5.20$ \\ 
$2.12$ & $-5.720$ & $-5.175$ & $5.273$ & $7.744$ & $5.20$ \\ 
%$2.55$ & $-5.779$ & $-5.768$ & $6.417$ & $9.426$ & $5.20$ \\ 
$3.00$ & $-5.834$ & $-6.315$ & $7.668$ & $11.261$ & $5.20$ \\ 
%$3.41$ & $-5.878$ & $-6.760$ & $8.879$ & $13.039$ & $5.20$ \\ 
%$4.17$ & $-5.950$ & $-7.476$ & $11.261$ & $16.542$ & $5.20$ \\ 
$4.92$ & $-6.012$ & $-8.102$ & $13.824$ & $20.305$ & $5.20$ \\ \hline
\end{tabular}
%\end{ruledtabular}
\end{center}
\caption{$(\Lambda =0)$ CDM: Time and space variations in $\protect\alpha $
for the corresponding redshifts of virialisation, $z_{v}$, in the $(\Lambda
=0)$ standard Cold Dark Matter model. The indexes $^{\prime }b^{\prime }$
and $^{\prime }c^{\prime }$, stand for background and cluster respectively.
The italic and bold entries correspond approximately to the Oklo and to the $%
\protect\beta $-decay rate constraint, respectively. The quasar absorption
spectra observations correspond to the values of $\Delta \protect\alpha /%
\protect\alpha |_{b}$. The initial conditions were set for $\protect\zeta /%
\protect\omega =-2\times 10^{-8}$, in order to have $\protect\alpha %
_{c}(z=0)=\protect\alpha _{0}$.}
\label{darkscdm}
\end{minipage}
\end{table*}

\begin{table*}%[tbp]
\begin{minipage}{150mm}
\begin{center}
%{\large{$w=-0.6$ CDM }}
\begin{tabular}{|c|c|c|c|c|c|c|}
\hline
$z_v$ & $\frac{\Delta\alpha}{\alpha} \vert_{b}\times 10^{-6}$ & $\frac{%
\Delta\alpha}{\alpha} \vert_{c}\times 10^{-7}$ & $\frac{\dot\alpha}{\alpha}
\vert_{b}\times 10^{-24}$ & $\frac{\dot\alpha}{\alpha} \vert_{c}\times
10^{-22}$ & $\frac{\delta\alpha}{\alpha}\times 10^{-6}$ \\ \hline
$0.00$ & $-5.440$ & $0.000$ & $0.332$ & $1.036$ & $5.44$ \\ 
$0.08$ & $-5.453$ & $-1.130$ & $0.385$ & $1.121$ & $5.34$ \\ 
$\mathit{0.17}$ & $\mathit{-5.466}$ & $\mathit{\ -2.077}$ & $\mathit{\ 0.440}
$ & $\mathit{\ 1.207}$ & $\mathit{\ 5.26}$ \\ 
$0.25$ & $-5.478$ & $-2.886$ & $0.497$ & $1.293$ & $5.19$ \\ 
$\mathbf{0.40}$ & $\mathbf{-5.502}$ & $\mathbf{\ -4.208}$ & $\mathbf{\ 0.613}
$ & $\mathbf{\ 1.469}$ & $\mathbf{\ 5.08}$ \\ 
%$0.48$ & $-5.513$ & $-4.759$ & $0.673$ & $1.558$ & $5.04$ \\ 
$0.50$ & $-5.517$ & $-4.914$ & $0.691$ & $1.586$ & $5.03$ \\ 
$1.08$ & $-5.599$ & $-7.828$ & $1.217$ & $2.354$ & $4.82$ \\ 
%$1.55$ & $-5.658$ & $-9.297$ & $1.704$ & $3.059$ & $4.73$ \\ 
$2.02$ & $-5.710$ & $-10.372$ & $2.238$ & $3.831$ & $4.67$ \\ 
%$2.62$ & $-5.770$ & $-11.415$ & $2.983$ & $4.909$ & $4.63$ \\ 
$3.02$ & $-5.806$ & $-11.980$ & $3.518$ & $5.684$ & $4.61$ \\ 
%$3.42$ & $-5.839$ & $-12.467$ & $4.076$ & $6.493$ & $4.59$ \\ 
%$4.20$ & $-5.898$ & $-13.275$ & $5.255$ & $8.207$ & $4.57$ \\ 
$4.59$ & $-5.924$ & $-13.626$ & $5.876$ & $9.107$ & $4.56$ \\ \hline
\end{tabular}%
\end{center}
\caption{$w=-0.6$ CDM: Time and space variations in $\protect\alpha $ for
the corresponding redshifts of virialisation, $z_{v}$, in the $w=-0.6$ CDM
model . The indices $^{\prime }b^{\prime }$ and $^{\prime }c^{\prime }$,
stand for background and cluster respectively. The italic and bold entries
correspond approximately to the Oklo and $\protect\beta $-decay rate
constraints, respectively. The quasar absorption spectra observations
correspond to the values of $\Delta \protect\alpha /\protect\alpha |_{b}$.
The initial conditions were set for $\protect\zeta /\protect\omega =-2\times
10^{-8}$, in order to have $\protect\alpha _{c}(z=0)=\protect\alpha _{0}$.}
\label{darkw6cdm}
\end{minipage}
\end{table*}

\begin{table*}%[tbp]
\begin{minipage}{150mm}
\begin{center}
%{\large{$w=-0.8$ CDM }}
\begin{tabular}{|c|c|c|c|c|c|c|}
\hline
$z_v$ & $\frac{\Delta\alpha}{\alpha} \vert_{b}\times 10^{-6}$ & $\frac{%
\Delta\alpha}{\alpha} \vert_{c}\times 10^{-7}$ & $\frac{\dot\alpha}{\alpha}
\vert_{b}\times 10^{-24}$ & $\frac{\dot\alpha}{\alpha} \vert_{c}\times
10^{-22}$ & $\frac{\delta\alpha}{\alpha}\times 10^{-6}$ \\ \hline
$0.00$ & $-5.412$ & $0.000$ & $0.340$ & $1.078$ & $5.41$ \\ 
$\mathit{0.09}$ & $\mathit{-5.427}$ & $\mathit{\ -1.587}$ & $\mathit{\ 0.401}
$ & $\mathit{\ 1.148}$ & $\mathit{\ 5.27}$ \\ 
$0.18$ & $-5.441$ & $-2.857$ & $0.462$ & $1.220$ & $5.16$ \\ 
%$0.26$ & $-5.455$ & $-3.899$ & $0.524$ & $1.294$ & $5.06$ \\ 
$0.34$ & $-5.468$ & $-4.777$ & $0.586$ & $1.370$ & $4.99$ \\ 
$\mathbf{0.42}$ & $\mathbf{-5.481}$ & $\mathbf{\ -5.527}$ & $\mathbf{\ 0.649}
$ & $\mathbf{\ 1.447}$ & $\mathbf{\ 4.93}$ \\ 
$0.52$ & $-5.498$ & $-6.362$ & $0.731$ & $1.549$ & $4.86$ \\ 
%$0.72$ & $-5.530$ & $-7.714$ & $0.905$ & $1.771$ & $4.76$ \\ 
$1.09$ & $-5.587$ & $-9.516$ & $1.260$ & $2.239$ & $4.64$ \\ 
%$1.55$ & $-5.650$ & $-10.975$ & $1.735$ & $2.884$ & $4.55$ \\ 
$2.01$ & $-5.706$ & $-11.997$ & $2.247$ & $3.595$ & $4.51$ \\ 
%$2.60$ & $-5.769$ & $-12.986$ & $2.976$ & $4.624$ & $4.47$ \\ 
$3.38$ & $-5.841$ & $-13.948$ & $4.023$ & $6.125$ & $4.45$ \\ 
%$4.15$ & $-5.901$ & $-14.689$ & $5.153$ & $7.757$ & $4.43$ \\ 
$4.92$ & $-5.953$ & $-15.297$ & $6.362$ & $9.510$ & $4.42$ \\ \hline
\end{tabular}%
\end{center}
\caption{$w=-0.8$ CDM: Time and space variations in $\protect\alpha $ for
the corresponding redshifts of virialisation, $z_{v},$ for the $w=-0.8$ CDM
model. The indices $^{\prime }b^{\prime }$ and $^{\prime }c^{\prime }$,
stand for background and cluster respectively. The italic and bold entries
correspond approximately to the Oklo and $\protect\beta $-decay rate
constraints, respectively. The quasar absorption spectra observations
correspond to the values of $\Delta \protect\alpha /\protect\alpha |_{b}$.
The initial conditions were set for $\protect\zeta /\protect\omega =-2\times
10^{-8}$, in order to have $\protect\alpha _{c}(z=0)=\protect\alpha _{0}$.}
\label{darkw8cdm}
\end{minipage}
\end{table*}

\begin{table*}%[tbp]
\begin{minipage}{150mm}
\begin{center}
%{\large{$\Lambda$ CDM }}
\begin{tabular}{|c|c|c|c|c|c|c|}
\hline
$z_v$ & $\frac{\Delta\alpha}{\alpha} \vert_{b}\times 10^{-6}$ & $\frac{%
\Delta\alpha}{\alpha} \vert_{c}\times 10^{-7}$ & $\frac{\dot\alpha}{\alpha}
\vert_{b}\times 10^{-24}$ & $\frac{\dot\alpha}{\alpha} \vert_{c}\times
10^{-22}$ & $\frac{\delta\alpha}{\alpha}\times 10^{-6}$ \\ \hline
$0.00$ & $-5.382$ & $0.000$ & $0.361$ & $1.068$ & $5.38$ \\ 
$\mathit{0.09}$ & $\mathit{-5.398}$ & $\mathit{\ -1.618}$ & $\mathit{\ 0.427}
$ & $\mathit{\ 1.128}$ & $\mathit{\ 5.24}$ \\ 
$0.18$ & $-5.414$ & $-2.889$ & $0.493$ & $1.193$ & $5.12$ \\ 
%$0.26$ & $-5.429$ & $-3.916$ & $0.558$ & $1.260$ & $5.04$ \\ 
$0.34$ & $-5.443$ & $-4.769$ & $0.623$ & $1.329$ & $4.97$ \\ 
$\mathbf{0.41}$ & $\mathbf{-5.458}$ & $\mathbf{-5.486}$ & $\mathbf{0.688}$ & 
$\mathbf{1.401}$ & $\mathbf{4.91}$ \\ 
$0.51$ & $-5.476$ & $-6.272$ & $0.773$ & $1.498$ & $4.85$ \\ 
%$0.61$ & $-5.493$ & $-6.943$ & $0.860$ & $1.602$ & $4.80$ \\ 
$1.07$ & $-5.573$ & $-9.141$ & $1.303$ & $2.161$ & $4.66$ \\ 
%$1.51$ & $-5.641$ & $-10.430$ & $1.773$ & $2.794$ & $4.60$ \\ 
$2.14$ & $-5.723$ & $-11.660$ & $2.508$ & $3.822$ & $4.56$ \\ 
%$2.54$ & $-5.769$ & $-12.239$ & $3.009$ & $4.536$ & $4.54$ \\ 
$3.31$ & $-5.844$ & $-13.132$ & $4.056$ & $6.045$ & $4.53$ \\ 
%$4.07$ & $-5.908$ & $-13.837$ & $5.191$ & $7.692$ & $4.52$ \\ 
$4.83$ & $-5.962$ & $-14.418$ & $6.408$ & $9.469$ & $4.52$ \\ \hline
\end{tabular}%
\end{center}
\caption{$\Lambda $ CDM: Time and space variations in $\protect\alpha $ for
the corresponding redshifts of virialisation, $z_{v}$, in the $\Lambda CDM$
model. The indices $^{\prime }b^{\prime }$ and $^{\prime }c^{\prime }$,
stand for background and cluster respectively. The italic and bold entries
correspond approximately to the Oklo and $\protect\beta $-decay rate
constraints, respectively. The quasar absorption spectra observations
correspond to the values of $\Delta \protect\alpha /\protect\alpha |_{b}$.
The initial conditions were set for $\protect\zeta /\protect\omega =-2\times
10^{-8}$, in order to have $\protect\alpha _{c}(z=0)=\protect\alpha _{0}$.}
\label{darklcdm}
\end{minipage}
\end{table*}

Variations in $\alpha $ also depend on the value and sign of the coupling, $%
\zeta /\omega $, of the scalar field responsible for variations in $\alpha $%
, to the matter fields. A higher value of $|\zeta /\omega |$, leads to a
stronger dependence on the density contrast of the matter inhomogeneities.
If the value or sign of $\zeta /\omega $ changes in space, then spatial
inhomogeneities in $\alpha $ occur. This could happen if we take into
account that on small enough scales, baryons will dominate the dark matter
density. The sign and value of $\zeta /\omega $ will change, and variations
in $\alpha $ will evolve differently on different scales. If there are no
variations in the sign and value of $\zeta /\omega $, then the only spatial
variations in $\alpha $ are the ones resulting from the dependence of the
fine structure 'constant' on the density contrast of the region in which one
is measuring $\alpha $. At first sight, one might conclude that the
difference between $\alpha _{c}$ and $\alpha _{b}$, is only a consequence of
the coupling of $\alpha $ to matter. In reality this is not so. It is always
possible to obtain the same results for any value $|\zeta /\omega |$ with a
suitable choice of initial conditions, as can be seen from the results in
Tables \ref{zeta1}, \ref{zeta4}, \ref{zeta6} and \ref{zeta10}.

The results of this paper offer a natural explanation for why any experiment
carried out on Earth \citep{fuj, prestage,prestage1}, or in our local solar
system \citep{olive}, gives constraints on possible time-variation in $%
\alpha $ that are much stronger than the magnitude of variation that is
consistent with the quasar observations \citep{murphylast} on extragalactic
scales. The value of the fine structure 'constant' on Earth, and most
probably in our local cluster, differs from that in the background universe
because of the different histories of these regions. It can be seen from the
Tables \ref{darkscdm}, \ref{darkw6cdm}, \ref{darkw8cdm} and \ref{darklcdm}
that inside a virialised overdensity we expect $\Delta \alpha /\alpha
\approx -10^{-7}$ for $z_{v}\leq 1,$ while in the background we have $\Delta
\alpha /\alpha \approx -10^{-6}$, independently of the structure formation
model used. The same conclusions arise independently of the absolute value
of the coupling $\zeta /\omega $, see Tables \ref{zeta1}, \ref{zeta4}, \ref%
{zeta6} and \ref{zeta10}.

The dependence of $\alpha $ on the matter-field perturbations is much less
important when one is studying effects of varying $\alpha $ on the early
universe, for example on the last scattering of the CMB or the course of
primordial nucleosynthesis \citep{martins,martins1,martins2}. In the linear
regime of the cosmological perturbations, small perturbations in $\alpha $
will decay or become constant in the radiation era \citep{mota2}. At the
redshift of last scattering, $z=1100$, it was found in \citep{mota3} that $%
\Delta \alpha /\alpha \leq 10^{-5}$. In the background, the fine structure
'constant', will be a constant during the radiation era \citep{bsbm} so long
as the kinetic term is negligible. A growth in value of $\alpha $ will occur
only in the matter-dominated era \citep{mota1}. Hence, the early-universe
constraints, coming from the CMB and primordial nucleosynthesis, are
comparatively weak, $\Delta \alpha /\alpha \leq 10^{-2}$, and are easily
satisfied.

\section*{Acknowledgments}

We would like to thank C. van de Bruck, M. Murphy, K. Subramanian and
J.K. Webb for
discussions. DFM was supported by Funda\c{c}\~{a}o para a Ci\^{e}ncia e a
Tecnologia, through the research grant BD/15981/98, and by Funda\c{c}\~{a}o
Calouste Gulb\^{e}nkian, through the research grant Proc.50096.

\appendix
\section{BSBM Equations}
\label{bsbmeq}
From the BSBM action (\ref{S}) the equation of motion for $\psi$ comes
\begin{equation}
\Box \psi = \frac{2}{\omega}e^{-2\psi}{\mathcal{L}}_{em}
\label{psiKG}
\end{equation}
The  right-hand-side (RHS) of equation (\ref{psiKG}) represents a
source term for $\psi$, which includes all the matter fields which are
coupled to it. These include not only relativistic matter (like
photons), but as well as
non-relativistic one that interact electromagnetically.
It is clear that ${\cal{L}}_{em}$,  vanishes for a sea of pure radiation
since  ${\cal{L}}_{em}=\left(E^2-B^2\right)/2=0$. The only significant
contribution to a variation in $\psi$ comes from nearly pure
electrostatic or magnetostatic energy associated to non-relativistic
particles. 
In order to make quantitative predictions we then need to know how much of the
non-relativistic matter contributes to the right-hand-side (RHS) of equation
(\ref{psiKG}). This can be parametrised by the ratio
$\zeta={\cal{L}}_{em}/\rho_m$, where $\rho_m$ is the energy density of the
non-relativistic matter. This non-relativistic matter, which  interact
electromagnetically, contributes to the Friedman equation
(\ref{fried1}) as $\rho_m\vert\zeta\vert e^{-2\psi}$.

For protons and neutrons,
$\zeta$ can be estimated from the electromagnetic corrections to the
nucleon mass, $0.63 MeV$ and $-0.13 Mev$, respectively
\citep{wep}. This correction contains the $E^2/2$ contribution, which
is always positive, and also term of the form $j_{\mu}a^{\mu}$, where
$j_{\mu}$ is the quarks' current \citep{bsbm}. Hence we take a guidance value of
$\zeta\approx10^{-4}$ for protons and neutrons. 

Using the parameter $\zeta$, the fraction of electric and magnetic
energies may then be written as \citep{havard}:
\begin{equation}
\zeta^{E}=\frac{E^2}{\rho_m} \qquad , \qquad \zeta^{B}=\frac{B^2}{\rho_m}
\label{zetas}
\end{equation}
where $E^2$ and $B^2$ are the electric and magnetic energies
respectively. Using (\ref{zetas}) in equation (\ref{psiKG}) we have \citep{havard}
\begin{equation}
\Box\psi=\frac{2}{\omega}e^{-2\psi}\rho_m \left(\zeta^{E}-\zeta^{B}\right)
\end{equation}
Since we are interested in the cosmological evolution of $\alpha$,
instead of using both parameters $\zeta^{E}$ and $\zeta^{B}$, we will 
use throughout this articles, the cosmological parameter, $\zeta$, defined as
$\zeta\equiv\zeta^{E}-\zeta^{B}$, which in the limit where
$\zeta^{E}\gg\zeta^{B}$ is positive, and when $\zeta^{E}\ll\zeta^{B}$
is negative.
Note that, the cosmological value of $\zeta$ has to be weighted, not only by the
electromagnetic-interacting baryon fraction, but also by the fraction of
matter that is non-baryonic. Hence the value and sign of $\zeta$ depends strongly on the
nature of dark matter to which the field $\psi$ might be
coupled. 

\section{The BSBM model and WEP violations}
\label{wepappendix}

In BSBM the test-particle Lagrangian may be split as $\mathcal{L}_{t}=%
\mathcal{L}_{m}+e^{-2\psi }\mathcal{L}_{em}$. Variation with respect to the
metric leads to a similar split of the stress-energy tensor, producing an
energy density of the form $\rho ((1-\zeta _{t})+\zeta _{t}e^{-2\psi }),$
and so a mass of $m((1-\zeta _{t})+\zeta _{t}e^{-2\psi }),$ (assuming
electric fields dominate). In order to preserve their ratios of $\zeta _{t}=%
\mathcal{L}_{em}/\rho $ test particles may thus be represented by \citep{wep4} 
\begin{equation}
\mathcal{L}(y)=-\int d\tau \;m((1-\zeta _{t})-\zeta _{t}e^{-2\psi })[-g_{\mu
\nu }\dot{x}^{\mu }\dot{x}^{\nu }]^{\frac{1}{2}}{\frac{\delta (x-y)}{\sqrt{-g%
}}}
\end{equation}%
where over-dots are derivatives with respect to the proper time $\tau $.
This leads to equations of motion: 
\begin{equation}
\ddot{x}^{\mu }+\Gamma _{\alpha \beta }^{\mu }\dot{x}^{\alpha }\dot{x}%
^{\beta }+{\frac{2\zeta _{t}e^{-2\psi }}{(1-\zeta _{t})-\zeta _{t}e^{-2\psi }%
}}\partial ^{\mu }\psi =0
\end{equation}%
which in the non-relativistic limit (with $\zeta _{t}\ll 1$) reduce to 
\begin{equation}
{\frac{d^{2}x^{i}}{dt^{2}}}=-\nabla _{i}\phi -2\zeta _{t}\nabla _{i}\psi \;,
\end{equation}%
where $\phi $ is the gravitational potential. Thus we predict an anomalous
acceleration: 
\begin{equation}
a={\frac{M_{s}}{r^{2}}}{\left( 1+{\frac{\zeta _{s}\zeta _{t}}{\omega \pi }}%
\right) }  \label{acc}
\end{equation}%
Violations of the WEP occur because $\zeta _{t}$ is substance dependent. For
two test bodies with $\zeta _{1}$ and $\zeta _{2}$ the E\"{o}tv\"{o}s
parameter is: 
\begin{equation}
\eta \equiv {\frac{2|a_{1}-a_{2}|}{a_{1}+a_{2}}}={\frac{\zeta _{s}|\zeta
_{1}-\zeta _{2}|}{\omega \pi }.}
\end{equation}%
This can be written more conveniently as the product of the following 3
factors \citep{bsbm}: 
\begin{equation}
\label{eta}
\eta =\left( {\frac{\zeta _{E}|\zeta _{1}-\zeta _{2}|}{\pi \zeta _{p}}}\
\right) \left( {\frac{\zeta _{p}}{\zeta}}\right) \left( {\frac{\zeta
}{\omega }}\right) .
\end{equation}%
where $E$ denotes Earth. If we take $\zeta _{n}\approx \zeta
_{p}\approx |\zeta _{p}-\zeta _{n}|=\mathcal{O}(10^{-4})$ then for typical
substances the first factor is $\approx 10^{-5}$. In order to satisfy
the existing  WEP violations experimental bounds \citep{will,wep3}, we need
to  ``play around'' with the two last terms of equation (\ref{eta}).
If we use the value assumed in \citep{bsbm}
$\zeta/\omega\approx-10^{-4}$.  Hence, we need $\zeta =%
\mathcal{O}(1)$ to produce $\eta =\mathcal{O}(10^{-13}),$ just an order of
magnitude below existing experimental bounds. If we
instead use $\zeta/\omega\approx-10^{-8}$ we would need  $\zeta=
\mathcal{O}(10^{-4})$ to satisfy the same  experimental bounds.
The
choice in the value of  $\zeta/\omega$ just depends on the nature of
dark-matter, but this is beyond the scope of this
paper.  Nevertheless, as shown in this paper,  any choice of
$\zeta_/\omega$ will give us the same cosmological behaviour for 
$\Delta\alpha/\alpha$ due to a degeneracy with the initial conditions
for $\psi$. Hence, the results
presented in this paper are not affected at all
by any experimental constraint imposed by WEP violations.

\label{lastpage}
\end{document}